\RequirePackage{fix-cm}

\documentclass[natbib,smallextended]{svjour3}

\usepackage{epsfig}                     
\usepackage{graphicx}                    
\usepackage{amssymb}                    
\usepackage{mathptmx}
\usepackage{latexsym}
\usepackage{color}                       
\usepackage{hyperref}   
\usepackage{breakurl}                 
\usepackage[normalem]{ulem}             
\usepackage{aas_macros}
\usepackage{natbib}

\newcommand{\lam}{$\lambda$}
\newcommand{\kms}{km~s$^{-1}$}
\newcommand{\ion}[2]{#1\,{\sc #2}}
\newcommand{\ecss}{erg~cm$^{-2}$~s$^{-1}$~sr$^{-1}$} 
\newcommand{\as}{$^{\prime\prime}$}

\def\tis{\!=\!}           
\def\tapprox{\!\approx\!} 
\def\HK{\mbox{H\,\&\,K}}
\def\hk{\mbox{h\,\&\,k}}
\def\Halpha{\mbox{H\hspace{0.1ex}$\alpha$}} 
\def\Lyalpha{\mbox{Ly\hspace{0.1ex}$\alpha$}}

\begin{document}



\title{Solar ultraviolet bursts}

%
\author{Peter R.\ Young \and Hui~Tian \and Hardi~Peter  \and Robert J.\ Rutten  \and Chris J. Nelson \and Zhenghua~Huang \and Brigitte Schmieder \and Gregal J. M. Vissers \and Shin Toriumi  \and Luc~H.~M.~Rouppe~van~der~Voort \and Maria~S.~Madjarska \and Sanja~Danilovic \and Arkadiusz~Berlicki \and L.~P.~Chitta \and Mark~C.~M.~Cheung \and Chad~Madsen \and Kevin~P.~Reardon \and Yukio~Katsukawa   \and Petr~Heinzel }

\institute{P. R.\ Young \at College of Science, George Mason University, 4400 University
  Drive, Fairfax, VA 22030, USA
  \and
  P. R.\ Young \at NASA Goddard Space Flight Center, Code 671,
  Greenbelt, MD 20771, USA
  \and 
  P. R.\ Young \at Northumbria University, Newcastle Upon Tyne NE1 8ST, UK
  \and
  H. Tian \at School of Earth and Space Sciences, Peking
   University, Beijing 100871, China
  \and 
  H. Peter \and M. S. Madjarska  \and L. P. Chitta \at Max Planck Institute for Solar System Research, Justus-von-Liebig-Weg 3, 37077, G\"ottingen, Germany
  \and
  R. J. Rutten \at Lingezicht Astrophysics, ’t Oosteneind 9, 4158 CA
   Deil, The Netherlands
\and
   R. J. Rutten \and L. H. M. Rouppe Van Der Voort   \at  Institute of Theoretical Astrophysics, University of Oslo, PO Box 1029 Blindern, NO-0315 Oslo, Norway
  \and
  C. J. Nelson \at School of Mathematics and Statistics, Hicks
  Building, University of Sheffield, Hounsfield Road, Sheffield S3
  7RH, UK
  \and
  C. J. Nelson \at Astrophysics Research Centre, School of Mathematics
   and Physics, Queen's University, Belfast, BT7 1NN Northern Ireland,
   UK
  \and
  Z. Huang \at Shandong Provincial Key Laboratory of Optical
  Astronomy and Solar-Terrestrial Environment, Institute of Space
  Sciences, Shandong University, Weihai, 264209 Shandong, China
  \and
  B. Schmieder \at LESIA, Observatoire de Paris, PSL Research
   University, CNRS, Sorbonne Universit\'es, UPMC Univ. Paris 06,
   Univ. Paris Diderot, Sorbonne Paris Cit\'e, 5 place Jules Janssen,
   92195 Meudon, France
  \and 
   G. J. M. Vissers \and S. Danilovic \at Institute for Solar Physics, Department of
   Astronomy, Stockholm University, AlbaNova University Centre, SE-106
   91 Stockholm, Sweden
  \and
  S. Toriumi \and Y. Katsukawa  \at National Astronomical Observatory of Japan, 
          2-21-1 Osawa, Mitaka, Tokyo 181-8588, Japan
\and
  L. H. M. Rouppe Van Der Voort  \at  Rosseland Centre for Solar Physics, PO Box 1029 Blindern, NO-0315 Oslo, Norway
  \and
  A. Berlicki \and P. Heinzel \at
   Astronomical Institute, Academy of Sciences of
   the Czech Republic, 25165 Ondrejov, Czech Republic
  \and 
  A. Berlicki \at Astronomical Institute, University of Wroc{\l}aw,
   Kopernika 11, 51-622 Wroc{\l}aw, Poland
  \and
   M. C. M. Cheung \at Lockheed Martin Solar and Astrophysics Lab, Palo
   Alto, CA 94304, USA
  \and
   C. Madsen \at Harvard-Smithsonian Center for Astrophysics, 60 Garden Street,
   Cambridge, MA 02138, USA
  \and
   C. Madsen \at Center for Space Physics, Boston University,
   725 Commonwealth Avenue, Boston, MA 02215, USA
  \and
  K. P. Reardon \at 2 National Solar Observatory, 3665 Discovery Dr.,
   Boulder, CO 80300, USA
  \and
  K. P. Reardon \at INAF-Osservatorio Astrofisico di Arcetri, Large
   Enrico Fermi 5, I-50125 Firenze, Italy
}

%

\maketitle

\begin{abstract}
The term ``ultraviolet (UV) burst'' is introduced to describe small, intense,
transient brightenings in ultraviolet images of solar active regions.
We inventorize their properties and provide a definition based on
image sequences in transition-region lines. 
Coronal signatures are rare, and most bursts are associated with small-scale, canceling opposite-polarity fields in the photosphere that occur in emerging flux regions, moving magnetic features in sunspot moats, and sunspot light bridges.
We also compare UV bursts with similar transition-region phenomena
found previously in solar ultraviolet spectrometry and with similar
phenomena at optical wavelengths, in particular Ellerman bombs. 
Akin to the latter, UV bursts are probably small-scale magnetic
reconnection events occurring in the low atmosphere, at photospheric
and/or chromospheric heights. 
Their intense emission in lines with optically thin formation gives
unique diagnostic opportunities for studying the physics of magnetic
reconnection in the low solar atmosphere. This paper is a review report from an International Space Science Institute team that met in 2016--2017.
\end{abstract}

%

\section{Introduction}

The Sun's chromosphere and corona were identified well before the
space age. 
The two orders of magnitude difference between their temperatures
suggested there is a thin, abrupt transition
in which the temperature rises very rapidly with height
\citep[e.g.,][]{1992str..book.....M}. 
Despite its thinness this ``transition region'' (TR) was found to emit
strongly in most resonance lines in the ultraviolet (UV) between 500
and 1600~\AA. 
The lines in this region are generally called ``TR lines'' and specific
phenomena observed in them ``TR phenomena''.
The lines are usually assigned characteristic formation temperatures in the
20--800\,kK range that are derived by assuming optically-thin ionization equilibrium. 
In this paper we will identify ions with their temperatures of maximum ionization ($T_{\rm max}$), but readers should be aware that for highly dynamic phenomena the emission may be formed in plasma out of equilibrium and the effective temperature of formation may differ from $T_{\rm max}$. This has been demonstrated in models of coronal loops by \citet{noci89} and in models  of the IRIS \ion{Si}{iv} and \ion{O}{iv} emission lines by \cite{2013A&A...557L...9D} and \citet{2013ApJ...767...43O}. Note that the high densities inferred for UV bursts \citep[e.g.,][]{2014Sci...346C.315P} may limit these effects as discussed by \citet{2018ApJ...857....5Y}.

Imaging in specific UV lines is rarely achieved due to technology
restrictions, so that so far most TR phenomena were classified through
their spectral UV features only, with particular emphasis on large Doppler
shifts, excessive broadening (larger than the thermal Doppler width
defined by the $T_{\rm max}$ value), and complex profiles.

The Interface Region Imaging Spectrometer \citep[IRIS,][]{iris} is the newest
UV spectrometer to observe solar TR lines.
The telescope feeds a slit spectrometer, providing spectra at unprecedented spatial and spectral resolution (Sect.~\ref{sect.previous}). In addition, a Slit-Jaw Imager (SJI) yields simultaneous images in four bandpasses. 
The one at 1400~\AA\ is of particular interest as it is often dominated by two \ion{Si}{iv} lines, making IRIS the first mission to routinely obtain high-resolution images in TR lines (Appendix~\ref{sect.imaging}). When combined with simultaneous spectroscopy of these lines, the environment of TR phenomena, including their history, can be identified, 
facilitating interpretation of their UV signatures in terms of other
solar fine structures found with other diagnostics.

In most energetic events, the two \ion{Si}{iv} line profiles have factor-two peak and profile ratios indicating optically thin line formation \citep{2014Sci...346C.315P, 2015ApJ...810...38K,2015ApJ...811L..33V}.
This is a valuable asset as their intensities therefore represent a
more direct measure of energy release than the complex, optically-thick,
source function mapping of lines such as the \ion{C}{ii} and
\ion{Mg}{ii} \hk\ resonance lines sampled by IRIS.  Their dominance of the IRIS 1400\,\AA\ slit-jaw passband over the
continuum contribution wherever there is substantial heating delivers
direct, high-cadence, wide-field imaging of such events (example in
panel b of Fig.~\ref{fig.sji}).

An early discovery from IRIS \citep{2014Sci...346C.315P}  was a type of very intense, compact
brightening in active regions with highly complex \ion{Si}{iv}
profiles called a bomb (``IRIS bombs'', IB) in view of
striking similarities with Ellerman bombs %
\citep[EB;][]{1917ApJ....46..298E}. 
This discovery inspired the first author to convene an International
Space Science Institute (ISSI) International Team to study these bombs
and other intense UV brightenings detected in IRIS spectra and images. 
The team met twice, in 2016 January and 2017 March, with the subject ``Solar UV bursts---a new insight to magnetic reconnection''. The term
\emph{UV burst} was chosen to denote the full scale of IRIS brightening events.  
Their discussions included the following major issues: How are UV
bursts related to EBs and the magnetic field? 
Can underlying patterns be recognised in different types?
Can complex burst line profiles be reproduced with numerical
simulation codes? 

The present article summarizes these discussions and the conclusions
reached at these meetings, and also provides observational criteria to
classify UV bursts in other IRIS data.

Section~\ref{sect.def} presents our definition of a UV burst.
Section~\ref{sect.previous} summarizes transient TR phenomena observed
with previous UV instrumentation. 
Section~\ref{sect.ibs} presents a summary of observational UV burst
results obtained so far from IRIS, while Section~\ref{sect.models}
describes efforts in modeling these. 
A final summary is given in Sect.~\ref{sect.summary}.

\section{UV burst definition}\label{sect.def}

Most solar phenomena are identified from their morphology in images or
image sequences. 
For example, relatively stable structures such as prominences, sunspots and
coronal loops are easily identified on individual images, while
transient features such as flares and jets are identified from
fast-cadence image sequences. 
Do UV bursts possess distinct signatures that can be readily employed for
their identification?

\begin{figure}[p]
\centerline{\epsfxsize=4.6in\epsfbox{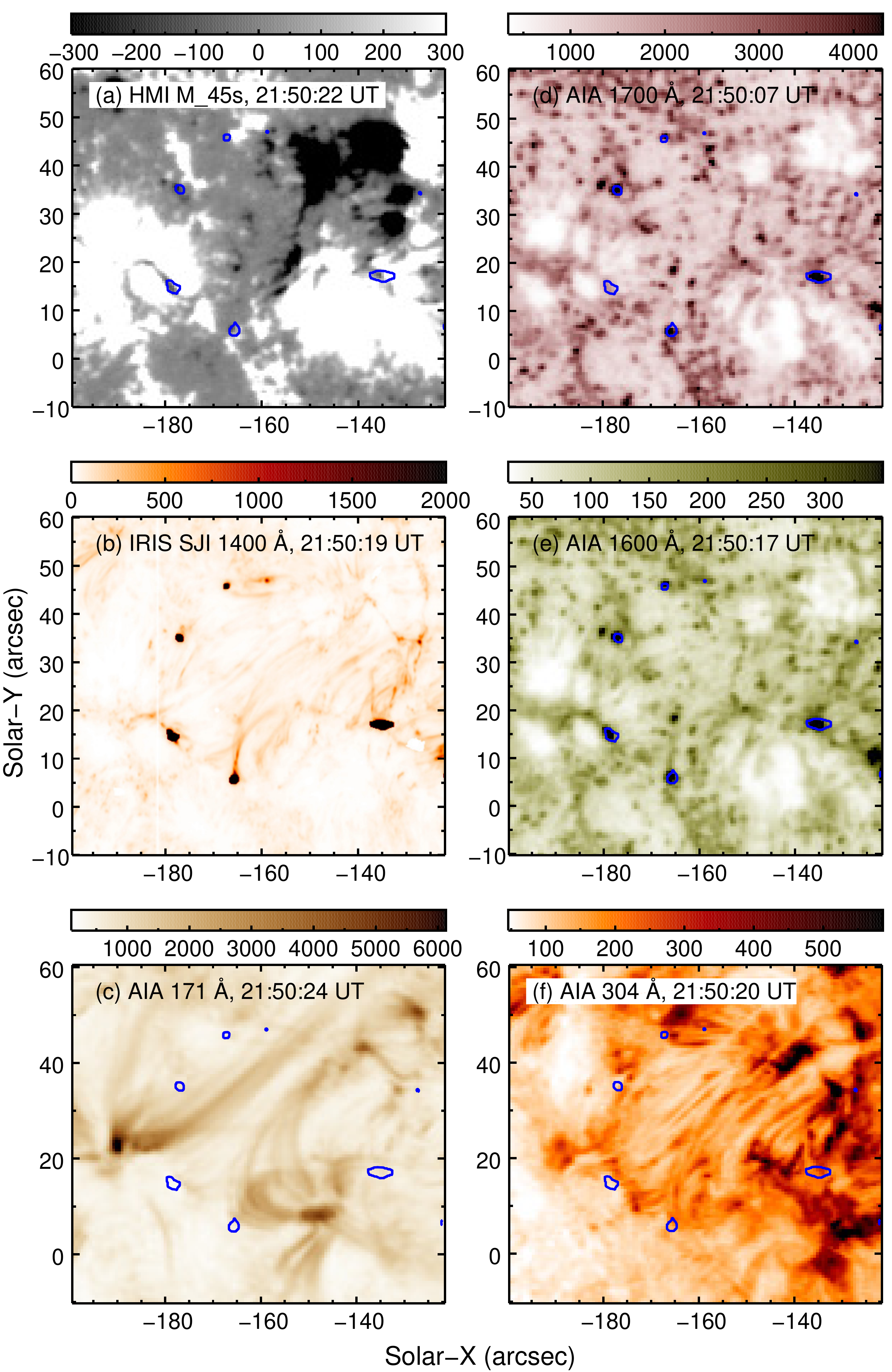}}
\caption[]{\label{fig.sji}
Images illustrating UV bursts from a 10-minute 18-second
cadence IRIS sequence showing part of active region AR 11875 on October 22,
2013 at solar $(X,Y) = (-162,25)$ during 21:45--21:55~UT. 
Panel (a): line-of-sight (LOS) HMI magnetogram scaled between $\pm 300$\,G.
Panel (b): IRIS SJI 1400\,\AA.
Panels (c)--(f): AIA 171\,\AA, 1700\,\AA, 1600\, \AA\ and 304\,\AA\ images.
Plotted quantity for panels (b)--(f) is intensity in DN~s$^{-1}$.
The blue contours superimposed on panels (a) and (c)--(f) outline the
brightest bursts in panel (b), at a level of 1000~DN~s$^{-1}$.
}
\end{figure}

Figure~\ref{fig.sji} shows near-simultaneous snapshots from 10-minute
active-region image sequences obtained with IRIS and the Solar
Dynamics Observatory (SDO).  
A movie version covering the whole sequence in the same format accompanies this manuscript.
The IRIS 1400~\AA\ slit-jaw image (SJI) in panel~(b) shows a number of
compact bright grains (black in the images) that are much
brighter than their surroundings. 
Viewing the full sequence in the movie shows that
their brightness flickers, that some are present for the entire
duration, and that others are short-lived. 
\emph{We call such features UV bursts.} 

Figure~\ref{fig.sji} also shows the co-temporal line-of-sight (LOS) magnetic
field from the Helioseismic and Magnetic Imager \citep[HMI;][]{hmi}  on board SDO, and
UV images from SDO's Atmospheric Imaging Assembly \citep[AIA;][]{aia}.
The AIA 171~\AA\ filter is dominated by \ion{Fe}{ix} \lam171.1, formed
at 0.8~MK.
The 304\,\AA\ filter is dominated by the \ion{He}{ii}
\Lyalpha-like resonance line and is formed around 80~kK.
The AIA 1600\,\AA\ and 1700\,\AA\ channels sample the upper
photosphere and lower chromosphere, with small magnetic concentrations appearing bright from their
evacuation.
If atmospheric heating is sufficiently large then the \ion{C}{iv}
\lam\lam 1548, 1550~\AA\
resonance lines can dominate over the continuum in
the 1600~\AA\ channel, and this is often the case for UV bursts hence
the stronger contrast compared to the 1700\,\AA\ channel. Note that
the 1600~\AA\ channel has a significantly broader filter width than
the SJI 1400~\AA\ channel and the continuum is more intense, thus the
SJI channel is much more effective for TR imaging.

Our name ``burst" denotes the brief duration of the brightening events.
As such it is a standard astronomy term, as in ``gamma-ray bursts''
and ``fast radio bursts''. 
This name has also been used before to describe solar TR phenomena
(Section~\ref{sect.bursts}).

We append ``UV'' to burst instead of ``transition region'' to keep
closer to observation and because ``region'' in the latter is somewhat
a misnomer. 
In the actual solar atmosphere every radial column must naturally
contain a steep temperature rise, but its height varies tremendously
both spatially and temporally and does not define a layer or shell as
in traditional one-dimensional equilibrium models such as the classical
ones of \citet{1981ApJS...45..635V}. 
A sharper and physically sounder description is to define the
chromosphere-corona transition at any location and instant as the
height where hydrogen reaches 50\% ionization---generally a
non-equilibrium quantity making up a highly warped, dynamic, and
history-dependent surface. 
There are probably no phenomena restricted solely to this
instantaneous transition surface; more likely heating events with
increasing ionization come from below (such as the bursts here)
whereas cooling events with increasing recombination come from above
(such as coronal rain). 
The ``TR'' phenomena reported in the older literature must
represent spectral UV signatures of wider-origin events. 
Our goal is to re-classify them into the latter by exploiting IRIS' UV
imaging capability. 

Thus, we prefer the term ``UV bursts''.
The UV wavelength range 465 to 1550~\AA\ (bounded by
\ion{Ne}{vii} \lam465 and \ion{C}{iv} \lam1550) is indeed dominated by
chromosphere-corona transition lines, although
there are also some chromospheric and coronal lines in this range.

Figure~\ref{fig.sji} and its movie version give a quick impression of
UV bursts; now we define them more formally. 
They are identified in image sequences sampling ultraviolet passbands
that are dominated by a chromosphere-corona transition emission line,
and have the following properties:

\begin{enumerate} \vspace*{-0.5ex} \itemsep=0.5ex

\item Compactness. 
Core brightenings $\lessapprox 2$\as\ in size, typically $\le 1$\as. 
A burst may appear with extended structure (jet, fibril, loop)
connected to it, but these are typically less bright. 
The burst itself may also appear spatially extended into one direction
(``flame''), 
but remains $\lessapprox 2$\as. 
Two or more bursts may appear close together but remain associated
with the same magnetic feature in the photosphere;

\item Duration. 
Bursts can have lifetimes ranging from tens of seconds to over an
hour. 
For long-lived ones the intensity is not constant, but may flicker
by about a factor two on timescales around a minute.
At high angular resolution such bursts can also appear as a sequence of
intermittent, repetitive flarings, possibly with a migrating footpoint;

\item Intensity. 
The burst is significantly brighter than the surroundings in the
pertinent UV passband.
For Figure~\ref{fig.sji}(b) the blue contours correspond to a factor
24 above the image median value. 
Peak SJI 1400~\AA\ intensities of individual spatial pixels of the
brighter events can reach factors of 100--1000 higher than the median.

\item Motion. 
UV bursts show only small proper motions, typically $\le 10$~\kms.
That is, they generally track dynamics of photospheric magnetic
features and their interactions, rather than traveling-front or wave motions
along extended structures such as loops;

\item UV bursts are not directly related to flares.
This condition distinguishes UV bursts from the compact, intense
kernels that appear along flare ribbons and otherwise look similar.

\end{enumerate}

We do not recommend setting a specific intensity threshold for UV
bursts, as there is no physical reason to exclude a wide spectrum of
events with different energies and temperatures defining the signal in
the pertinent wavelength bands. 
Instead, a dataset-specific threshold suits better to let the observer
obtain a manageable number of events for study, for example by
specifying a factor above the median intensity value (such as 24 for
panel~b of Figure~\ref{fig.sji}), or a factor k$\sigma$ above the
median value with $\sigma$ the standard deviation of the
intensity over an image ($k\tapprox 7$ for the 1400\,\AA\ data in
Figure~\ref{fig.sji}). 
Also note that integrated intensity over the burst area may define
a more appropriate threshold than the intensity of the brightest
spatial pixel in the burst.

The above criteria also include most IBs identified in the literature,
but with the important difference that the latter are defined through
their spectroscopic signatures discussed in Section~\ref{sect.ibs}. 
At least one type of event, the narrow line bursts of
\citet{2016ApJ...829L..30H}, 
do not satisfy the IB criteria while satisfying the UV burst criteria. 
Therefore we envision IBs to be a subclass of UV bursts, although
probably the dominant one.

Section~\ref{sect.ibs} describes key results from the IRIS papers
in more detail; we summarize these here and we refer the reader to Section~\ref{sect.ibs} for further details and references to the literature:
\begin{itemize} \vspace*{-0.5ex} \itemsep=0.5ex

\item \ion{Si}{iv} line profiles of UV bursts mostly have a complex
shape that can include multiple peaks, large excess broadening beyond the thermal width (at temperature $T_{\rm max}$),
and very extended wings.  
They often indicate the presence of bi-modal jets; in slanted limbward
viewing these may also be mapped into profile differences along
different LOSs to the burst.
However, some events show simple, Gaussian-shaped profiles not much
wider than the thermal width;

\item UV bursts generally overlie or are directly adjacent to small
magnetic features in the photosphere that interact (usually
cancelation following convergent proper motion) with similar or larger
opposite-polarity features;

\item Most UV bursts do not show significant co-spatial brightening in
the AIA 171~\AA\ channel (dominated by \ion{Fe}{ix}, formed at 0.8~MK), nor in the other coronal AIA
channels, and only rarely in the \ion{He}{ii} 304\,\AA\ passband;

\item EBs are more common than UV bursts, with around 10--20\%\ of EBs
having a burst signature while 30--60\% of bursts show an EB
signature. 

\end{itemize}
These should be considered properties of UV bursts, but are not part
of their definition above. 

The general consensus from the literature and our ISSI team is that UV
bursts are small-scale magnetic reconnection events, and that their
complex \ion{Si}{iv} line profiles possibly display dynamics associated with a
current sheet; for example fast-moving, dense plasmoids\footnote{The term plasmoid applies to magnetic islands that appear in models of current sheets subject to the tearing instability and  also to observed features in the Earth's magnetotail, where the magnetic field can be measured in situ. For remote-sensing solar observations, the term is often applied to small-scale brightenings, but it is not possible to determine if these are the same as the current sheet/magnetotail plasmoids without magnetic field measurements.}, or turbulence
induced in surrounding plasma by reconnection outflow jets. 
The reconnection occurs somewhere in the upper photosphere-to-upper
chromosphere regime, rather deep in the atmosphere.
The particular location and also the line of sight to the reconnection site may be responsible
for differences in burst signatures such as variations in \ion{Si}{iv}
profiles.

Our UV burst definition requires transition region imaging capability,
which was either not available with instruments prior to IRIS or
infrequently used (Appendix~\ref{sect.imaging}). 
If a slit spectrometer is raster-stepping to build an image scan, only
infrequent snapshots of the burst will be obtained, and just one if it is
short-lived. 
If there is only one snapshot the evolution and proper motion cannot
be analyzed. 
However, it may be possible to estimate at least the latter from
supporting data, for example, magnetograms (HMI or better) or
magnetic-concentration monitoring in AIA 1600 and 1700~\AA\ images.
Also, if the burst intensity happens to be low at the time(s) of
snapshot sampling it may fail high-intensity threshold criteria. 
Therefore raster mode data tend to underestimate the 
number
of UV
bursts, but some may get identified properly.

In sit-and-stare mode following mean solar rotation (10\,\as/hr
at disk center) the time evolution can be accurately monitored as long
as the burst remains within the slit, but if the burst has low
intensity during this sampling period it may not be flagged at all.
Also, the spatial extent of a burst cannot be established with
fixed-slit data, which may lead to erroneous feature classification.

\section{Previous event types}\label{sect.previous}

\begin{table}[t]
\caption{Properties of TR instruments.}\label{tbl.inst}
\begin{tabular}{lcccc}     
\hline
\noalign{\smallskip}
&&Wavelength & \multicolumn{2}{c}{Resolution} \\
\cline{4-5}
\noalign{\smallskip}
Instrument &Years & Range/\AA & Spatial/arcsec & Spectral/\AA\\
\hline
\noalign{\smallskip}
Skylab S082B & 1973--1974 & 970--1940 & 2 &0.06 \\
HRTS & 1975--1992 & 1170--1700 & 1 & 0.1\\
SMM/UVSP & 1980--1989 & 1070--3600 & 4 & 0.05\\
SOHO/SUMER & 1996--2017 & 500--1600 & 2--3 & 0.04\\
SOHO/CDS & 1996--2013 & 308--381, 515-630 & 6--10 & 0.3--0.5 \\
Hinode/EIS & 2006--present & 170--212, 246--292 & 3--4 & 0.06 \\
IRIS & 2013--present & 1332--1358, 1389--1407 & 0.3 &  0.026 \\
\hline
\end{tabular}
\end{table}

TR event types are generally quite different to coronal events such as coronal hole plumes, active region loops and coronal bright points. These have typical spatial extents of 10's of arcseconds and lifetimes from hours to days, compared to sizes of a few arcseconds or less, and lifetimes of minutes  for TR events. In addition the availability of coronal EUV and X-ray imaging means coronal features are typically identified from spatial morphology. 
The various TR event types have almost exclusively been identified through
UV
spectrometry, and so definitions  to some extent depend on the capabilities of the
instruments, in particular their spatial resolution, spectral
resolution and coverage. 
Table~\ref{tbl.inst} summarizes the characteristics of the instruments
treated in this section. 
Note that due to multiple configurations or changes in time, the
listed parameters are not complete; the listed values are typical for
TR measurements by these instruments. The table shows that IRIS has better spatial and spectral resolutions compared to other instruments by factors of three and two, respectively.

The first report of strongly broadened TR lines was
\citet{1976SoPh...47..127B} 
who presented spectra of TR instabilities observed as very broad
features in the \ion{C}{iv} \lam\lam1548, 1550 doublet observed with
the \emph{Skylab} S082B instrument. 
This spectrometer had effectively no spatial resolution along the
slit, so that further analysis awaited flights of the High Resolution
Telescope Spectrograph (HRTS) rocket experiment.
HRTS was the first solar instrument to observe the TR at high spectral
and spatial resolution. 
It was flown ten times on rockets between 1975 and 1997 and once as
part of the Spacelab 2 payload in 1985\footnote{A brief history of
HRTS is available at \url{http://wwwsolar.nrl.navy.mil/hrts\_hist5.html}.}.

Long-term monitoring at high spatial resolution of the TR began with
the UltraViolet Spectrometer and Polarimeter (UVSP) on board the Solar
Maximum Mission during 1984 to 1989. 
The Solar Ultraviolet Measurements of Emitted Radiation (SUMER) and
Coronal Diagnostic Spectrometer (CDS) on board the Solar and
Heliospheric Observatory (SOHO) were the next major spectrometers with TR
coverage, observing for almost 20 years since 1996. 
The EUV Imaging Spectrometer (EIS) on board the Hinode spacecraft was
launched in 2006 and, although mostly focused on the corona, it
observes a number of upper TR emission lines.
Finally, IRIS was launched in 2013.  It observes UV lines 
with significantly higher spatial and spectral resolution
than previous instruments, and adds highly-valuable slitjaw imaging.

We now discuss different types of TR features that were identified
with these instruments in the light of our UV bursts.  
The emission lines that were most commonly used are given in
Table~\ref{tbl.lines}, with  $T_{\rm max}$ values derived from the CHIANTI database \citep{chianti1,chianti8,2016JPhB...49g4009Y}.
We suggest that especially the so-called SMM bursts
(Section~\ref{sect.bursts}) and possibly also active region blinkers
(Section~\ref{sect.blinkers}) directly correspond to UV bursts, based
on the emission measure comparison in Appendix~\ref{sect.em}.

\begin{table}[t]
\caption{Key UV lines sampling the chromosphere-corona transition.}\label{tbl.lines}
\begin{tabular}{ccll}     
\hline
Ion & $\log\,(T_{\rm max}/{\rm K})$& Wavelength (\AA) & Instruments  \\  
\hline
\ion{Si}{iv} &4.90 & 1393.78, 1402.77 & HRTS, UVSP, SUMER, IRIS   \\
\ion{C}{iv} & 5.05 & 1548.19, 1550.77 & HRTS, UVSP, SUMER \\
\ion{O}{iv} & 5.15 & 1401.16 & HRTS, UVSP, SUMER, IRIS \\
\ion{N}{v} & 5.30 & 1238.82, 1242.80 & HRTS, SUMER \\
\ion{O}{v}  &5.40 & 629.73 &  CDS, SUMER(x2) \\
            && 192.90 & EIS \\
\hline
\end{tabular}
\end{table}

\subsection{Jets}

In recent solar physics the term jet is usually applied to transient,
elongated and extending structures that are seen in emission in image sequences. The term surge pre-dates the use of jet in the solar physics literature \citep[e.g.,][]{1966soat.book.....Z} and is generally used to describe similar structures that are mostly seen in absorption in chromospheric lines (particularly H$\alpha$). 

The lack of TR imaging capability prior to IRIS means that the analogs of coronal or chromospheric jets could not be directly identified. Instead TR spectra have been reported of jets identified from other imaging data. For example, \citet{1983A&A...127..337S} studied  SMM/UVSP spectra of a surge identified from H$\alpha$ images, and they measured blueshifts and line broadening in the \ion{C}{iv} lines. Similarly, \citet{madjarska11} captured SOHO/SUMER spectra of a coronal jet identified from EUV and X-ray imaging data, and found blueshifts of up to 300~\kms\ in several TR lines.

The term jet was applied  specifically to a TR spectral feature by \citet{1983ApJ...272..329B} who identified intense emission in
\ion{Si}{iv} and \ion{C}{iv} lines that always appeared blue-shifted and
extended out to 400~\kms. Subsequent work showed that these features are quite rare, however \citep{1994SSRv...70...21D}.

Since the launch of IRIS, jets can now be identified directly from TR imaging data, and examples include the quiet Sun network jets of \citet{2014Sci...346A.315T} and the active region jets of \citet{2015ApJ...801...83C}.

A jet may correspond to a UV burst if there is an intense brightening
at the jet base that passes the criteria of Section~\ref{sect.def}. 
The column of the jet, where large velocities are expected, generally
would not pass the criteria because the emission is typically weak and
extends over several or more arcseconds.

\subsection{Penumbral microjets and sunspot dots}

Penumbral microjets were first described by
\citet{2007Sci...318.1594K} based on Hinode Solar Optical Telescope
filtergram images centered on the \ion{Ca}{ii} H line. 
They are small jets occurring within a sunspot's penumbra with lengths
typically 1 to 4~Mm and lifetimes $\le$1~minute. 
Many can be identified at any one time within a penumbra.
They often have a different orientation than the adjacent penumbral
filaments. 
Cospatial observations with IRIS and the Swedish Solar Telescope (SST)
in \ion{Ca}{ii}
\citep{2015ApJ...811L..33V} 
demonstrated that these features also appear in \ion{Si}{iv}
as a compact brightening at the top of the jet with an
intensity enhancement of around five compared to the surroundings.

\citet{tian14-dots} 
found brightness features in IRIS SJI 1330~\AA\ and 1400~\AA\ images
that also occur mainly in sunspot penumbrae, which they termed
\emph{bright dots}. 
They have a progressively weaker signal in lower atmosphere layers, with most events having no signature in H$\alpha$ \citep{2016ApJ...829..103D}.
\citet{tian14-dots} suggested that bright dots have a connection to the \ion{Ca}{ii} penumbral microjets
which was largely confirmed by
\citet{2015ApJ...811L..33V}, 
but \citet{2017ApJ...835L..19S} 
identified a distinct class of bright dots without \ion{Ca}{ii}
signature.
Penumbral bright dots were also found in 193~\AA\ images from the Hi-C
rocket flight \citep{2016ApJ...822...35A}, 
which were not considered to be coronal features but contributed by
cooler TR emission lines in the 193~\AA\ passband. \citet{2016ApJ...816...92T} found signatures of larger penumbral jets in the AIA coronal filters, and suggested a different formation mechanism for these compared to the standard microjets.
Bright dots in a sunspot umbra were identified by
\citet{2016A&A...587A..20C}; 
they lie at the footpoints of coronal loops.

We note that sunspot bright dots are excluded from our UV burst
definition due to their relatively weak intensity enhancement in SJI
1400~\AA\ images and their large proper motions of 10--40~\kms\
\citep{tian14-dots}. 
However, since penumbral microjets and some bright dots are found at
locations of convective intrusion into strong field, the physics of
these events may have similarities to the bursts found in umbral light
bridges (Section~\ref{sect.mag}).

\subsection{Explosive events}\label{sect.ee}

This name was introduced by %
\citet{dere89} 
for events found with the third HRTS rocket flight in 1979; earlier
they were named ``turbulent events'' by %
\citet{1983ApJ...272..329B}. 
They were identified as large broadening of \ion{C}{iv} \lam1548, 
on one side of the line only or on both sides but with
asymmetrical wings.
Wherever both wings were enhanced there was often a spatial
separation along the slit of up to 2\as. 
The maximum Doppler shift was not found to vary with the
position of the events on the solar disk, suggesting that the
corresponding mass motions are isotropic. 
This led the authors to regard the events as explosions, hence
the name ``explosive event''.

Most of the HRTS-era explosive events were measured in quiet areas,
with perhaps the most significant active region event reported by
\citet{1988ApJ...335..986B} with the HRTS on Spacelab~2 in 1985. 
This was reported to be only five times more intense than nearby
active region plage. 
Although difference in spatial resolution must be accounted for when
comparing intensities of compact features from different instruments,
this relatively low enhancement suggests that the various HRTS flights
did not observe anything directly equivalent to the IBs of
\citet{2014Sci...346C.315P} 
since those are characterized by much larger intensity enhancements of
the \ion{Si}{iv} lines.

There were a number explosive event studies from SUMER data
\citep[see][for a summary of results]{huang14}, but almost all were
focused on quiet Sun or coronal holes as active regions on the disk
were rarely observed due to concerns over instrument degradation. 
One exception is the event presented by \citet{2000AdSpR..26..457B}
that showed high velocity features in ions ranging from \ion{N}{iv} to
\ion{Ne}{viii} ($\log\,T_{\rm max}({\rm K})=5.2$--5.8) although, based on
the images presented in the paper, it did not show the strong
intensity enhancement characteristic of IBs.

The only IRIS paper that has presented a quiet Sun explosive event is that of \citet{huang14}. The event occurred at the boundary of an equatorial extension of a polar coronal hole, and it exhibited strongly broadened \ion{Si}{iv} lines. The high spatial resolution of the SJI 1330~\AA\ images allowed small jets to be seen, which are the first direct evidence of the jets inferred from the high Doppler broadening 
in spectral data. \citet{2015ApJ...809...82G} and \citet{2017MNRAS.464.1753H} studied active region events that they referred to as explosive events, but we consider these to be UV bursts and they are discussed in the following sections.


\subsection{Bursts}\label{sect.bursts}

The use of the word ``burst" with regard to TR phenomena dates back to
the \emph{Skylab} period, with \citet{emslie78} and
\citet{1982ApJ...258..835W} referring to impulsive solar bursts. 
These showed sudden increases in UV emission-line intensities, with
two events associated with flares and two not. 
Of direct relevance to the IRIS observations is the study by
\citet{hayes87} who reported observations from the SMM/UVSP instrument
with the \ion{Si}{iv} \lam1402.77 and \ion{O}{iv} \lam1401.16 lines
that are also observed by IRIS. 
They defined bursts as events for which the \ion{Si}{iv} \lam1402.77
intensity increased by a factor two in one minute. 
The line width and velocity were not part of this definition. 
The emission measure analysis presented in Appendix~\ref{sect.em}
suggests that they observed equivalents of IBs.

We also note that \cite{1997SoPh..175..341I} used the term burst to
refer to a contiguous train of explosive events, lasting for up to
30~minutes.  Studies of some IRIS bursts suggest similar behavior
\citep[e.g.,][]{2015ApJ...812...11V,2015ApJ...809...82G}.

\subsection{Blinkers}\label{sect.blinkers}

The term blinker was introduced by \citet{1997SoPh..175..467H} to
describe small-scale brightenings seen in quiet-Sun areas with the
Coronal Diagnostic Spectrometer (CDS) on board SOHO. 
More detailed studies followed by \citet{bewsher02} and
\citet{parnell02}, who applied an automatic detection method.
They were identified from peaks in light curves for individual
emission lines by requiring that the peak be a factor $P$ times the
noise level of nearby light-curve minima. 
In addition, this criterion must be satisfied for $N$ neighboring
spatial pixels. 
Ranges of $P$ and $N$ values were investigated, with $P \tis 5$ and $N \tis 3$
chosen for statistical analysis.
The \ion{O}{v} \lam629.7 line was the strongest TR line observed by
CDS and was the main reference line used in these studies.

Unlike SUMER, the CDS instrument routinely observed active regions on
the solar disk so, although it lacked the spectral resolution
required to study broadened profiles of explosive events or IBs,
it was well capable of detecting large UV-line enhancements.
An early study by \citet{young97} found two intense UV brightenings in
the core of a small, recently-emerged active region that were factors
of $\approx$~50 brighter than the average quiet Sun. 
The study of active region blinkers of \citet{parnell02} yielded a
wide spectrum of blinker events, with the most intense being around a
factor 100 times stronger than average active-region emission. 
The emission measure analysis of Appendix~\ref{sect.em} shows that
\ion{O}{v} blinkers are about an order of magnitude weaker than UV
bursts in \ion{Si}{iv}, but we consider them to be consistent with the
UV burst definition.
\citet{young04} studied a handful of the most intense active region
blinker events and found densities as high as
$10^{11} - 10^{12}$~cm$^{-3}$, although only one event exhibited
significantly enhanced non-thermal broadening.

The Hinode/EIS instrument mostly observes coronal emission lines, but
a number of lines with $T_{\rm max}$ values from 0.1 to 0.7~MK
are also observed, as highlighted by \citet{young07-tr}, although they are significantly weaker than the strong TR lines observed by CDS, SUMER and IRIS.
An intense UV brightening was identified in  \citet{young07-tr} and considered
analogous to CDS active-region blinkers. 
Despite this, we are aware of only one other comparable brightening
reported in the literature, that of
\citet{2010ApJ...724.1083G}. 
A systematic search for these brightenings in existing EIS data-sets
and additional joint EIS--IRIS observing campaigns are required to
determine if there is any correspondence between the EIS events and IBs.


\subsection{Ellerman bombs}\label{sect.eb}

\citet{1917ApJ....46..298E} referred to these as ``solar hydrogen bombs'', but
Ellerman bomb (EB) is the preferred name today.
EBs are photospheric features with sizes $\le 2$\as\ and lifetimes of
several minutes that have traditionally been identified only in
complex emerging active regions, but similar phenomena occur in
sunspot moats
and were also recently found in quiet-Sun areas
\citep{2016A&A...592A.100R, 2017ApJ...845...16N}. 
They are defined by their signature in the H$\alpha$ line, consisting
of strong enhancements in the red and blue wings. 
Crucially, these wing intensity enhancements are significantly larger
than for the much more common quiescent magnetic flux concentrations
that cause wing brightening by the hot-wall effect, a phenomenon
different from EBs but sometimes causing confusion
\citep{2013JPhCS.440a2007R}. 
Several studies
\citep{vissers13,2013ApJ...779..125N,2015ApJ...812...11V,2015ApJ...798...19N,2015ApJ...805...64R,2016ApJ...823..110R}
suggest that enhancement of about 50\%\ above the mean at $\pm 1$~\AA\ from
\Halpha\ center is a good criterion to separate EBs from flux
concentrations.
When observed towards the limb at high angular resolution (at least of
0.2\as) EBs are found to have distinctive flame-like appearance in
images obtained in the wing of H$\alpha$ \citep{2010PASJ...62..879H,2011ApJ...736...71W}.
Recent 3D MHD simulations have been able to reproduce this tell-tale
characteristic remarkably well \citep{2017A&A...601A.122D,
2017ApJ...839...22H}.

EBs occur where bi-polar small-scale magnetic fields move together and
cancel, typically at locations of emerging flux in active regions as
stipulated by \citet{1917ApJ....46..298E}, but also at moving magnetic features
(MMF) in moats around sunspots.
They are interpreted as magnetic reconnection events occurring below
1000~km height
\citep{2002ApJ...575..506G,2004ApJ...614.1099P,2011ApJ...736...71W,vissers13}.
Emerging-flux EBs were discussed in the review article of
\citet{2015sac..book..227S}.

EBs are sometimes found at the base of \Halpha\ surges---large jet-like
structures seen in absorption in filtergrams taken in a \Halpha\ wing---
but less often than claimed by \citet{1973SoPh...28...95R}. 
He also reported surges from EB-like brightenings in light bridges.
These different magnetic environments are further discussed in
Section~\ref{sect.mag}.

\section{UV burst results from IRIS}\label{sect.ibs}

Most IRIS studies in this context focused on IRIS bombs (IBs) that
were introduced by \citet{2014Sci...346C.315P} who named these for similarities with
Ellerman bombs. 
Other terms have also been used, such as hot explosions
\citep{2015ApJ...810...38K}, explosive events
\citep{2015ApJ...809...82G,2017MNRAS.464.1753H}, compact brightenings \citep{2016A&A...593A..32G} and flaring
active-region filaments \citep{2015ApJ...812...11V}.
In the present discussion we assume that they are all UV bursts under the
definition in Section~\ref{sect.def} above.

\begin{figure}[p]
\centerline{\epsfxsize=\textwidth\epsfbox{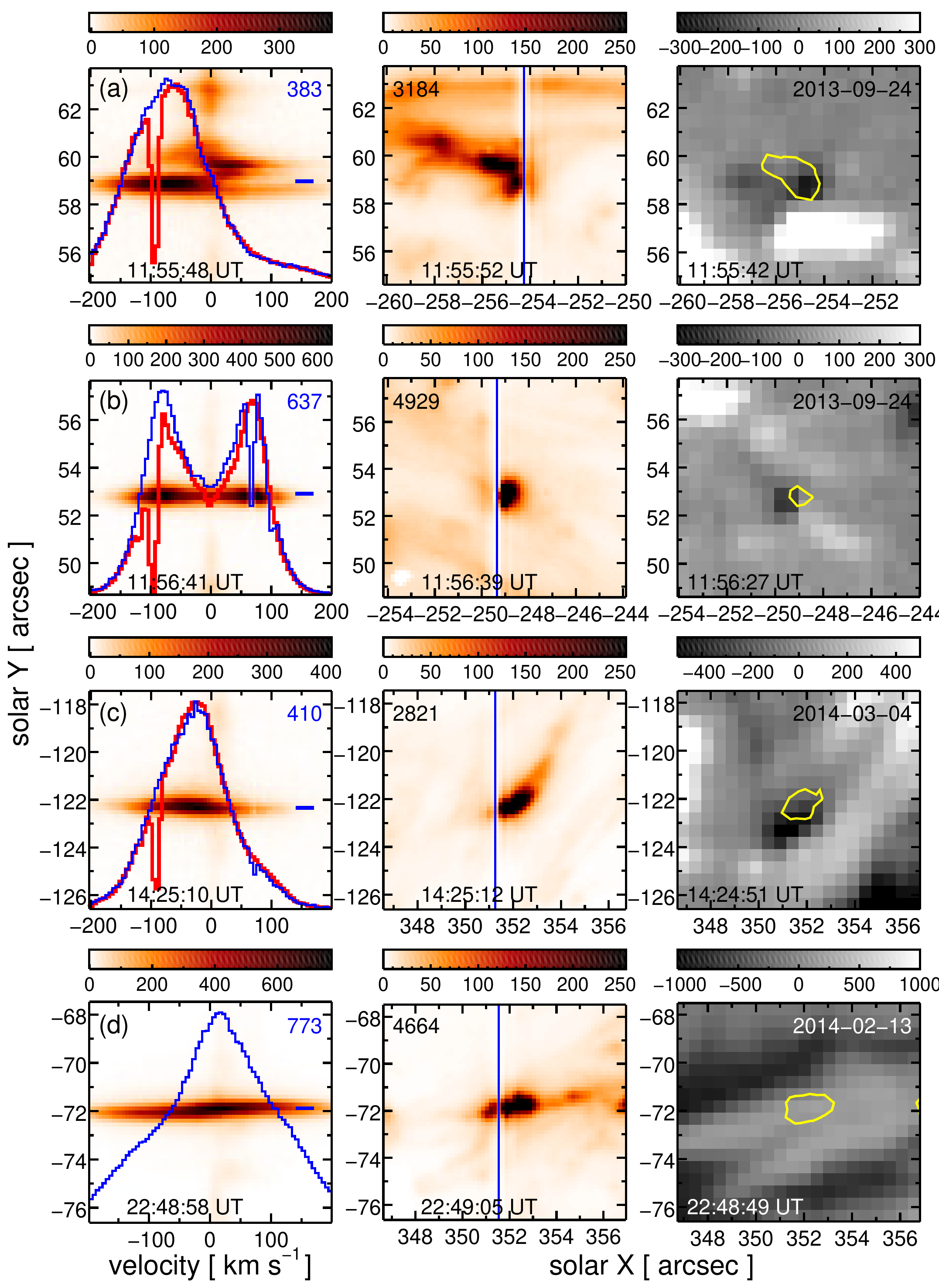}}
\caption[]{Four examples of intense bursts. 
The leftmost panels show IRIS spectrograms for the spectral band
$\pm 200$~\kms\ around \ion{Si}{iv} \lam1402.77; the center panels
show SJI 1400~\AA\ images; the rightmost panels show HMI LOS
magnetograms. 
The blue and red curves superimposed in the first column show the 1D spectra for \ion{Si}{iv} \lam1402.77 and \lam1393.76, respectively, at the
$Y$-pixel specified by the short blue line at right, with \lam1393.76 scaled down by a factor two.
The yellow contours in the third column are taken from corresponding AIA
1600~\AA\ images.  The blue vertical lines in the center panels
show the position of the IRIS slit. 
The blue numbers in the left panels give the maximum specific intensity
in units $10^3$~\ecss~\AA$^{-1}$  for the 1D spectra; the color bar gives the same units. The numbers in the center panels give the maximum image intensity in DN.
The observation
dates are shown at top-right of the rightmost panels.}
\label{fig.bursts}
\end{figure}

To illustrate the properties of IBs and UV bursts in this section, 
we show four examples from the literature in
Figure~\ref{fig.bursts}. 
Bursts (a) and (b) correspond to bomb events 4 and 1, respectively, of
\citet{2014Sci...346C.315P}; burst (c) is the event studied by
\citet{2015ApJ...809...82G}; burst (d) is an event from the data-set
studied by \citet{2015ApJ...811..137T}. 
For each burst we show the spectral image for \ion{Si}{iv} \lam1402.77
(leftmost panels), with the nearest-in-time IRIS slit-jaw (SJI)
1400~\AA\ image (middle panels) and the corresponding 
LOS magnetograms from HMI (rightmost panels).
The images were spatially aligned using the AIA 1600~\AA\ channel
images, which also reveal the bursts (yellow contours in right panels). 
We emphasize the high intensity of the \ion{Si}{iv} lines by
displaying values of the lines' maximum specific intensities in the left panels. 
For comparison, an average quiet Sun \ion{Si}{iv} \lam1402.77 profile has a peak specific intensity of about 800~\ecss~\AA$^{-1}$ and a full-width at half-maximum of 0.2~\AA. Average active region line profiles typically have peaks in the 5000--20\,000~\ecss~\AA$^{-1}$ range.

The IBs were identified by \citet{2014Sci...346C.315P} from a $140\times170$~arcsec$^2$ IRIS raster scan
taken on 2013 September 24 during 11:44--12:04~UT.     
The authors did not explicitly state criteria that defined IBs, but they 
formulated the four properties given below. We consider criteria (2) to be too strict, as noted in the following text.

\begin{enumerate} \vspace*{-0.5ex} \itemsep=0.5ex
\item the \ion{Si}{iv} lines are very wide with wings
reaching out to $\approx 200$~\kms\ separation from line center;

\item the \ion{Si}{iv} intensities are enhanced by a factor
$\sim$1000 compared to the surrounding active region; 

\item narrow atomic and single-ionized absorption lines are 
superimposed as absorption blends on the IRIS lines, including the
\ion{Si}{iv} lines;

\item the ratio of \ion{Si}{iv} \lam1402.77 to \ion{O}{iv} \lam1401.16
becomes much larger than usual for areas where both appear, suggesting
very high densities $\gtrsim 10^{13}$~cm$^{-3}$.

\end{enumerate}

Whereas very wide lines are typical of explosive events
(Sect.~\ref{sect.ee}), properties (2)--(4) were not previously
reported for such events.
\citet{2015ApJ...811...48Y} demonstrated that recognizing property
(3) requires the high spectral resolution of IRIS.

The line width criterion (1) is required to ensure the line is  broad enough 
to display the narrow absorption blends. Note that it does not refer to the full-width at half-maximum of the line, as is often used for describing Gaussian-shaped lines.
The strongest absorption line (\ion{Ni}{ii} \lam1393.33) lies at $-91$~\kms\ from the
center of \ion{Si}{iv} \lam1393.76---see the red profiles in the left panels of Figure~\ref{fig.bursts}(a)--(c). 

The IB intensity criterion above applies only to the most extreme events, and  many of the IBs in the literature do not reach this intensity. Also, since IBs are defined from their spectroscopic signature, the intensity at the instant the IRIS slit crosses the event is unlikely to be the highest value. This is particularly relevant when trying to associate IBs with features identified from imaging data
\citep[e.g.,][]{2017ApJ...836...52Z}. 
\citet{2015ApJ...809...82G} demonstrated that IB criteria (1), (3) and
(4)
are valid for most of the lifetime of the event they studied,
which was captured with a sit-and-stare observation. 
If only a single snapshot had been obtained with a raster scan, then
the event may have failed criterion (2). 
Assessing the feature intensity by measurement on
IRIS SJI 1400~\AA\ images is to be preferred. 
It has not yet been demonstrated whether events exist that satisfy IB criteria
(1), (3) and (4) but have only a weak intensity enhancement, say a
factor 5 to 10.

To put the IB intensities in context, we first use an IRIS quiet Sun
dataset from 2013 October 3  04:20--04:37~UT to measure an average quiet
Sun \ion{Si}{iv} \lam1402.77 intensity of 160~\ecss.
The average intensity\footnote{\citet{2014Sci...346C.315P} did not give
intensities in calibrated units; the values quoted in this
sentence were obtained by the lead author of the present study.} of the
active region studied by \citet{2014Sci...346C.315P} was 2800~\ecss, and the
intensity of the brightest spatial pixel of IB number~1 of
\citet{2014Sci...346C.315P} was $5.2\times 10^5$~\ecss.  
We caution that the sensitivity of IRIS has degraded significantly
since launch, with version 4 of the radiometric calibration (as
implemented through the IDL procedure \verb|iris_get_response|) showing a
decrease in sensitivity of
a factor 4 at 1402.8~\AA\ between 2013 August and 2017 August.
It is therefore recommended to always quote intensities in calibrated
units rather than data number (DN) units in order that bursts at
different times can be compared.

\begin{figure}[t]
\centerline{\epsfxsize=\textwidth\epsfbox{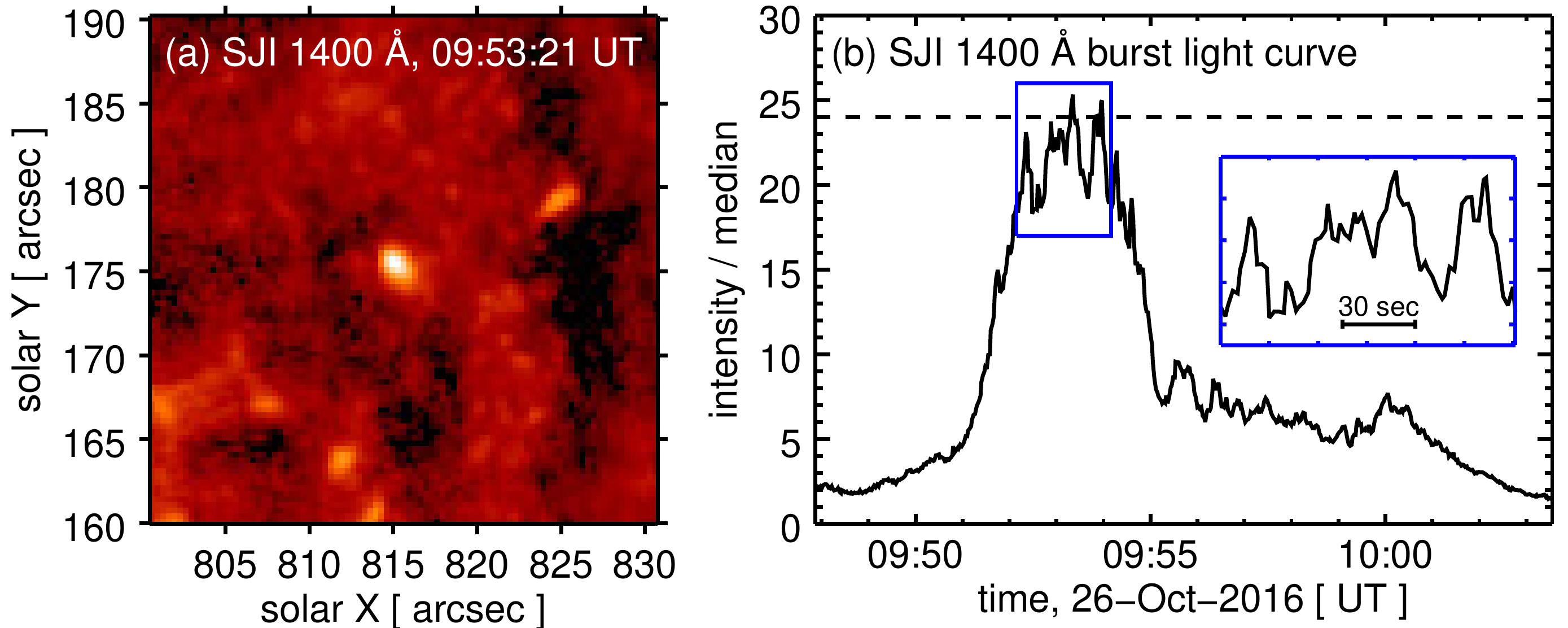}}
\caption[]{Rapid variability in a UV burst. Panel (a) shows an IRIS SJI 1400~\AA\ image of a burst (image center) with a logarithmic intensity scale. Panel (b) shows the intensity variability of this burst from an area of $7\times 8$ pixels centered on the feature and expressed relative to the median of the larger area displayed in panel (a). The blue box indicates the region shown in greater detail in the inset plot. The horizontal dashed line shows the intensity level used for the IRIS intensity contours in Figure~\ref{fig.sji}.}
\label{fig.lc}
\end{figure}

One feature of UV bursts is the
flickering of their intensities over time. 
Light curves constructed from the IRIS/SJI 1400~\AA\ images are shown
by \citet{2014Sci...346C.315P}, \citet{2015ApJ...812...11V} and
\citet{2015ApJ...810...38K} and reveal that the intensity evolution of
a UV burst consists of many individual peaks that are often close to
the SJI temporal resolution.
The cadence per SJI channel is usually lower than for the spectrometer exposures due to cycling between different SJI channels. 
The fastest possible cadence of SJI imaging is 1.7~seconds, and we have identified one data-set beginning at 09:08~UT on 2016 October 16 that has this cadence for the SJI 1400~\AA\ channel and contains bursts. Figure~\ref{fig.lc}(a) shows an image of one burst from this dataset, captured at the peak brightness. Panel (b) shows the light curve of the burst and the inset plot shows a close up of the variability near the peak intensity. Individual spikes at the sequence cadence can be identified, which suggests that the full variability of the burst is not being captured at this cadence. However, we also note that the frame-to-frame intensity variation for this event is relatively small at around 10\%\ or less.

\citet{2015ApJ...809...82G} presented sit-and-stare observations that captured  a UV burst with a five second cadence. Figure~7 of this work shows that the  \ion{Si}{iv} emission line light curve exhibits strong variability on this timescale.
These authors applied wavelet analysis to identify periods of 30 and
60--90~seconds in the \ion{C}{ii} and \ion{Si}{iv} spectral lines.
We note that flickering has previously been reported for EBs, for
example by \citet{2007A&A...473..279P} and
\citet{2011ApJ...736...71W}. 

\subsection{Emission line profiles}\label{sect.prof}

The \ion{Si}{iv} profiles of IBs presented in the literature show
significant shape variations; Figure~\ref{fig.bursts} gives some
flavor of this. 
The striking double-peaked profile found by \citet{2014Sci...346C.315P} and shown
in Figure~\ref{fig.bursts}b is not common, but has been seen in other
events, for example, Figure~3 of \citet{2017ApJ...851L...6R} and
Figure~7 of \citet{2016ApJ...824...96T}. 
More commonly the line has a dominant blue or red component
(panels a and c of Figure~\ref{fig.bursts}) or only a
small Doppler shift of the centroid (Figure~\ref{fig.bursts}d). 
Some profiles show \ion{Si}{iv} wings extending to over 2.5~\AA\
(530~\kms) from line center \citep[Figure~16
of][]{2015ApJ...812...11V}, and a common feature is a triangular
profile shape, i.e., the profiles' sides are linear when plotted as
wavelength vs.\ log(intensity) as opposed to the parabolic shape of
Gaussian profiles. 
Examples are shown in Figures~4 to 8 of \citet{2016ApJ...824...96T}. 

\begin{figure}[t]
\centerline{\epsfxsize=0.9\textwidth\epsfbox{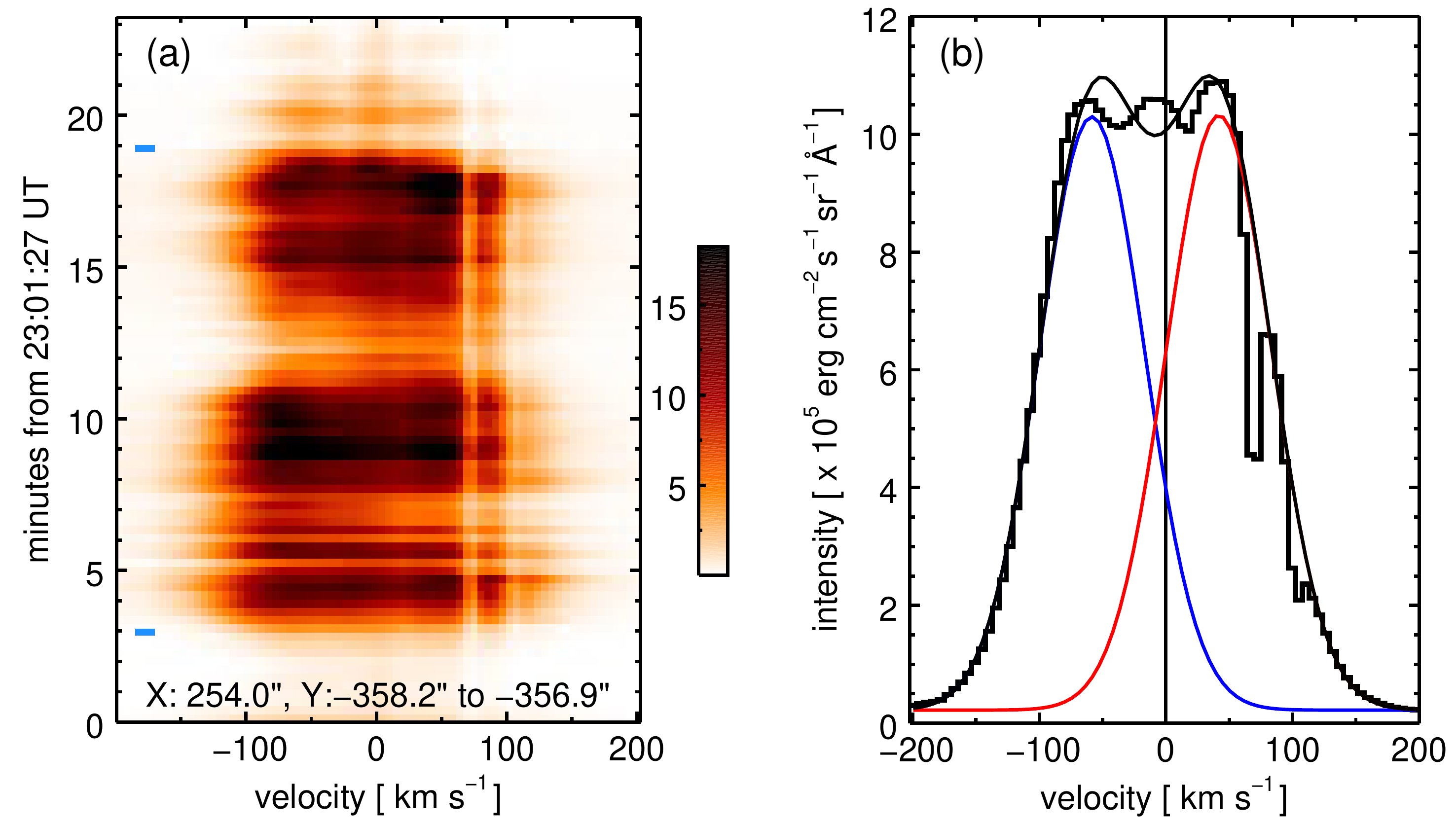}}
\caption{Panel (a) shows the evolution of the line profile of \ion{Si}{iv} \lam1402.8 for a UV burst observed on 2014 October 24. The spatial coordinates of the burst are indicated and a reverse, linear intensity scaling is applied. Short horizontal blue lines identify the time range that was averaged to yield the line profile (thick black line) in panel (b). Two Gaussians were fit to the profile and are shown as red and blue lines; the thin black line shows the total fitted profile.}\label{fig.time}
\end{figure}

One possibility for the complex shapes of IB line profiles is that the IRIS exposures represent a time average of a set of simpler, but rapidly-varying line profiles. For example, one could imagine that individual plasmoids are being rapidly produced by the plasma, each with a characteristic large speed but in different directions. The sum of the line profiles would then yield a broad, complex spectral feature. The observations in the literature typically use exposure times of 2 to 8~seconds so the characteristic timescale for the evolution of individual line profiles  would have to be a fraction of a second. Evidence against this comes from the widths of the emission lines in sit-and-stare sequences, which can remain approximately the same over timescales of minutes. 
An  example is the evolution of the \ion{Si}{iv} lines in the UV burst presented in Figure~2 of  \citet{2015ApJ...809...82G}. Although the line widths vary significantly with time, they remain approximately constant for periods of about a minute (about 20 exposures).  Another example is shown in Figure~\ref{fig.time} where the evolution of \ion{Si}{iv} \lam1402.8 for a burst observed on 2014 October 24 at 23:02~UT is shown, and we see that the line width remains about 200~\kms\ for 15~minutes. This is difficult to explain unless the basic velocity structure of the plasma remains fairly stable over timescales of minutes.

The \ion{Si}{iv} profiles often vary significantly across IBs along
the slit, for example in Figure~2 of \citet{2015ApJ...811...48Y},
Figure~6 of \citet{2016A&A...593A..32G}, and Figure~5 of
\citet{2017A&A...605A..49C}. 
In the latter case a MMF was studied and redshifts were found on the
forward side of the MMF, blueshifts on the rearward side, which was
interpreted as reconnection jets tilted with respect to the line of
sight.

A defining feature of IBs is the presence of atomic
or single-ionized  blends superimposed on the broad \ion{Si}{iv} emission lines. 
The deepest of the absorption lines is usually \ion{Ni}{ii} \lam1393.33, superimposed on \ion{Si}{iv} \lam1393.76.  
Sometimes the latter is not returned in the IRIS telemetry stream, however, and  absorption is less easy to identify in \ion{Si}{iv} \lam1402.77 (compare profiles in the leftmost column of Figure~\ref{fig.bursts}).
When present, the  narrow absorption dips have relatively
small Doppler shifts (generally less than 10~\kms) and betray the presence of
relatively cool gas (hydrogen and helium predominantly neutral)
along the line of sight of the burst that produces the wide profiles. 

Occasionally absorption features near the rest velocities of the \ion{Si}{iv} lines can be identified, and examples have been shown by \citet{2015ApJ...811...48Y} and \citet{2015ApJ...812...11V}. We note that the latter paper suggested the absorption occurs in an overlying layer rather than due to the burst emission itself being optically thick.

\begin{figure}[t]
\centerline{\epsfxsize=\textwidth\epsfbox{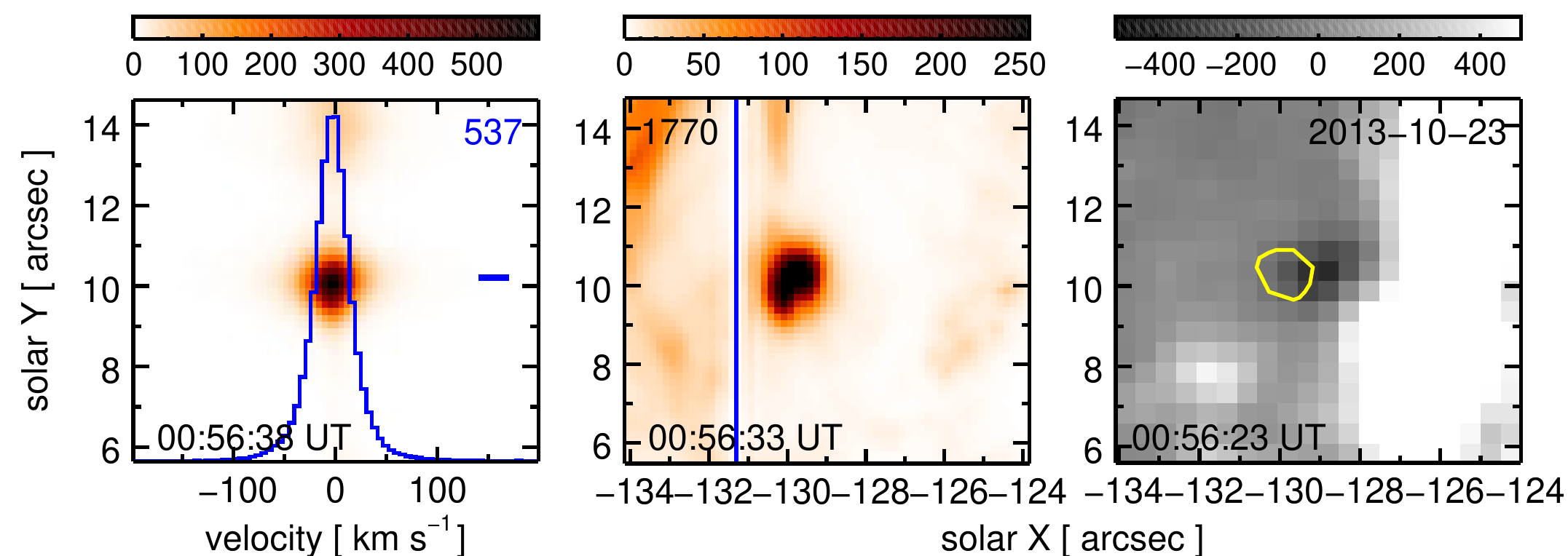}}
\caption[]{An example of a narrow-line UV burst from an  IRIS raster obtained between 00:41 and 00:51~UT on 2013 October 23. The format is the same as for Figure~\ref{fig.bursts}.}
\label{fig.narrow}
\end{figure}

In addition to the complex line profile patterns, bursts with very
narrow, Gaussian-shaped \ion{Si}{iv} line profiles have been reported
by \citet{2016ApJ...829L..30H}. They were found above sunspots and the authors suggested a connection with sunspot plumes.   
The most intense event had an intensity of $3.5\times 10^5$~\ecss\ in
the brightest spatial pixel, comparable to the IBs of
\citet{2014Sci...346C.315P}.
These events thus appear to satisfy the UV burst criteria, but the
narrow line profiles disqualifies them from being IBs. 
The ISSI team found other examples of intense, narrow-line
bursts, and Figure~\ref{fig.narrow} shows an example from 2013 October 23, although this one is not above a sunspot. 
A key requirement is to distinguish narrow-line bursts from mini-flare
ribbons (Section~\ref{sect.mini}) and dynamic loops
(Section~\ref{sect.loops}), that may also show narrow line profiles. 
A wider survey of narrow-line bursts and a study of their underlying
magnetic field structure will be worthwhile.

The complex UV burst profiles suggests that some details of the
reconnection physics are being revealed in the UV lines; a first
modeling effort was performed by \citet{2015ApJ...813...86I} who
considered how plasmoids in a current sheet may lead to line
profile shapes as observed, in particular triangular
profiles. 
New high spatial resolution chromospheric images from the CHROMIS
instrument \citep{2017ApJ...851L...6R} show tiny brightenings of sizes close to 0.1\as\ that are associated with a burst and may correspond to plasmoids. The authors also simulated \ion{Si}{iv} emission using a 2.5D radiative MHD simulation and found that plasmoids could be responsible for the complex \ion{Si}{iv} profiles if the line-of-sight  passes through multiple plasmoids along the current sheet.

\subsection{Energy estimates for UV bursts}

In this section we give a rough estimate of the energy of  a typical burst, using the example from 2014 October 24 shown in Figure~\ref{fig.time}. We assume that the burst plasma is isothermal at the temperature of formation of \ion{Si}{iv} (80~kK). As noted in the previous section the \ion{Si}{iv} profile remains approximately the same during the event lifetime (Figure~\ref{fig.time}) and we average the \lam1402.8 line over the 16~minute period indicated in Figure~\ref{fig.time}a to yield the profile shown in panel b. For simplicity we fit this with two Gaussians drawn as red and blue lines on Figure~\ref{fig.time}b. The fit does not reproduce the fine structure at the top of the profile, but the width and total intensity are accurately reproduced.
The Gaussians have LOS velocities of $-58$ and $+42$~\kms, equal intensities and widths of $4.9\times 10^5$~\ecss\ and 0.46~\AA, respectively. The latter corresponds to a non-thermal velocity of 59~\kms\ after subtracting the thermal and instrumental widths. The burst image drifted through the IRIS slit during the observation, and the full lifetime derived from SJI images was 36~minutes. 
The burst is about 1~arcsec (725~km) in size. The \ion{O}{iv} \lam1401.2 emission line is very weak and we estimate a \ion{Si}{iv} \lam1402.8/\ion{O}{iv} \lam1401.2 ratio of 317, corresponding to an electron number density of $6.3\times 10^{12}$~cm$^{-3}$ using the results of \citet{2018ApJ...857....5Y}.


Given the large Doppler velocities of the two line components and the small size of the burst, the event essentially blows itself apart in about 7~seconds (the travel time from the center of the burst to the edge). The burst can thus be considered to regenerate itself 309 times during the 36~minute lifetime. We refer to these as sub-bursts.

For each sub-burst, the energy inputs are as follows. The particles need to be heated from 10 to 80~kK; they are given an instantaneous kinetic energy corresponding to a bulk speed  of $\approx 50$~\kms; they are given random non-thermal velocities of 59~\kms; and heating is applied during the sub-burst lifetime to balance the radiative losses and maintain the burst's temperature.

The number of particles in the burst at any one time is, on average, given by $n=2.3EM_V/N_{\rm e}$, where the volume emission measure is defined in Appendix~\ref{sect.em}. The average \ion{Si}{iv} intensity is $9.8\times 10^5$~\ecss, thus giving $EM_V=2.9\times 10^{46}$~cm$^{-3}$ and $n=1.1\times 10^{34}$. If we assume these particles are heated from 10 to 80~kK, then the energy required for a sub-burst is $1.5nk\Delta T=1.6\times 10^{23}$~erg.

The combined bulk flow and random motions are given by $0.5 nm(v^2 +\xi^2)$, where $m$ is the average mass of a particle taken here as 0.6 the mass of a proton. This gives a kinetic  energy of $3.3\times 10^{23}$~erg for each sub-burst.

The energy loss rate due to radiation is given by $EM_V Q(T)$, where $Q(T)$ is the radiative loss function. At 80~kK it takes the value of $4\times 10^{-22}$~erg~cm$^3$~s$^{-1}$  calculated using CHIANTI with photospheric abundances. For the 7~seconds lifetime of the sub-burst, the radiative losses are $8.1\times 10^{25}$~erg, over two orders of magnitude larger than the thermal and kinetic energy terms, 

Multiplying these energy estimates by the number of sub-bursts gives the total energy requirement of $2.5\times 10^{28}$~erg for the burst's lifetime of 36~minutes. This compares with the value $5\times 10^{28}$~erg derived with a different method for the IB studied by \citet{2014Sci...346C.315P}. We also note that  a typical energy for a X-class flare is $10^{32}$~erg \citep[e.g.,][]{2005JGRA..11011103E}.

\subsection{Magnetic environments and signatures}\label{sect.mag}

The UV bursts reported in the literature are generally characterized
by rapid evolution of small-scale magnetic elements on the
photospheric surface, often evidenced in the 45~second cadence LOS
magnetograms from HMI but sometimes requiring better resolution and
sensitivity. 
Three types of magnetic environment are recognized to harbor UV bursts:

\begin{enumerate} \vspace{-0.5ex} \itemsep=0.5ex

\item Emerging flux regions (EFRs) in complex active regions.
Small flux elements in these display fast streaming motions;
cancellation occurs regularly.

\item Moving magnetic features (MMF) in sunspot moats.
These can have opposite polarity to the spot in sea-serpent
patterns \citep{1973SoPh...28...61H} 
with cancellation against same-polarity features.

\item Light bridges (LB). 
Elongated features with highly-sheared magnetic field that cross
sunspot umbrae, or occur close to them \citep{2003A&ARv..11..153S}.

\end{enumerate}

EFR examples include the events in \citet{2014Sci...346C.315P},
\citet{2015ApJ...812...11V}, \citet{2017ApJ...836...63T,
2017ApJ...836...52Z} and
\citet{2017ApJ...851L...6R}. 
\citet{2015ApJ...809...82G} and \citet{2017A&A...605A..49C} studied
MMF events. 
Bursts associated with an LB were discussed by
\citet{2015ApJ...811..137T} and \citet{2018ApJ...854...92T}.

HMI magnetograms are shown in the rightmost column of
Figure~\ref{fig.bursts}. 
Events (a)--(c) are located within 1\as\ of small negative polarity
features with vertical flux densities of around 200--400~G. 
Event (d) from \citet{2015ApJ...811..137T} is located to the north
side of a light bridge running east-west and does not show an obvious
compact magnetic feature, but the authors noted a continuous supply of
field with LOS flux densities of 200--400~G within the light
bridge that kept driving brightenings. 
Hinode Spectropolarimeter data showed that this field was highly
inclined with horizontal flux densities of around 1000~G. 
This highlights the importance to obtain the full vector magnetic
field when studying UV burst evolution.

Light bridges are believed to be convective intrusions into strong
background magnetic field
and have mostly horizontal field \citep{1997ApJ...484..900L}. 
Reconnection between this field and the surrounding vertical field
generates EB-like phenomena 
and surges of cool plasma \citep{1973SoPh...28...95R}. 
Recent, high-resolution H$\alpha$ data from the SST show that the surges resolve into a ``fan" of many jets side-by-side \citep{2016A&A...590A..57R}.
IRIS data have demonstrated that UV bursts are rooted in LBs at the
bases of the surges
\citep{2015ApJ...811..138T,2017ApJ...848L...9H,2018ApJ...854...92T}.
Since reconnection events are constantly driven by the convection in the
LB, repeated bursts and surges can be sustained for hours but
individual events are short-lived \citep{2001ApJ...555L..65A}.

\begin{figure}[t]
\begin{minipage}{0.6\textwidth}
\includegraphics[width=2.7in]{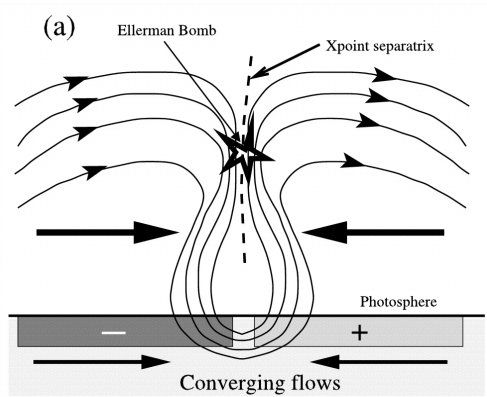}
\end{minipage}\hfill
\begin{minipage}{0.4\textwidth}
\caption{A cartoon from \citet{2002ApJ...575..506G} indicating how an Ellerman bomb can be triggered by converging horizontal flows in the photosphere. Reproduced by permission of the AAS.}\label{fig.georgoulis}
\end{minipage}
\end{figure}

\begin{figure}[t]
\begin{minipage}{0.6\textwidth}
\includegraphics[width=2.7in]{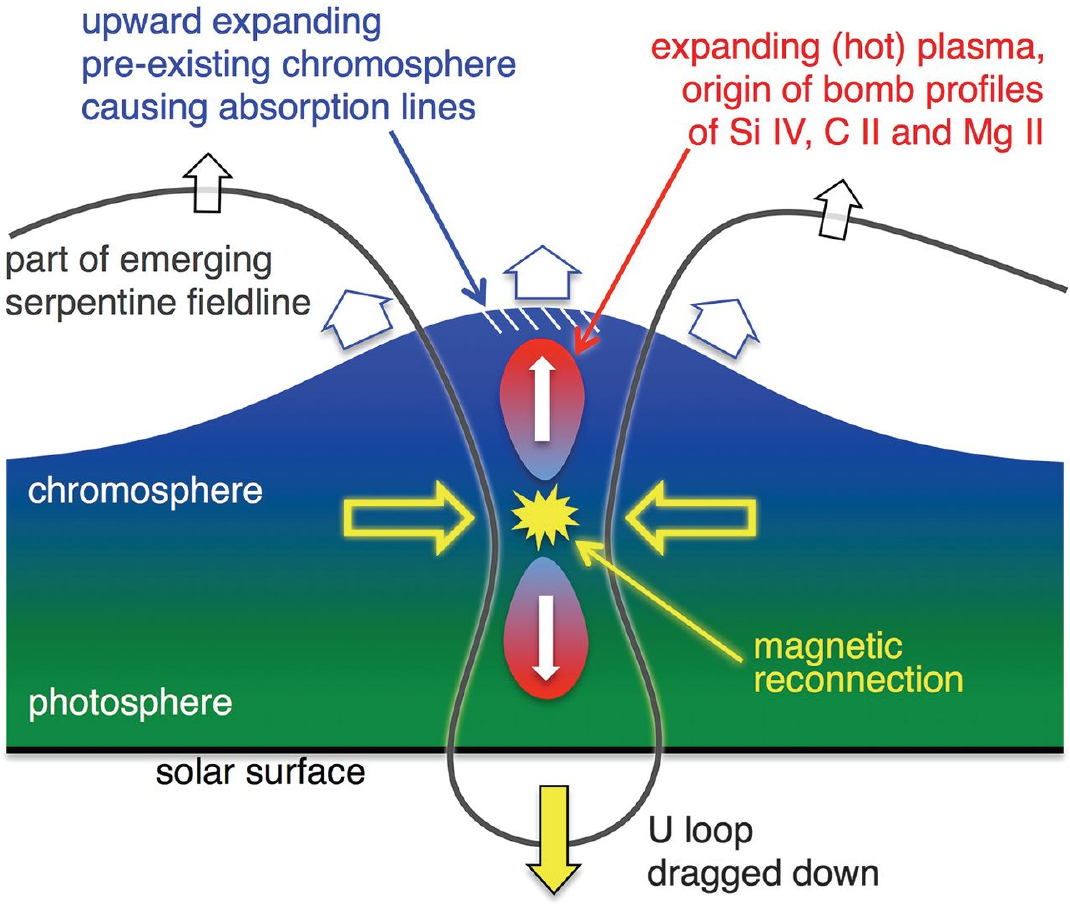}
\end{minipage}\hfill
\begin{minipage}{0.4\textwidth}
\caption{A cartoon for the formation of an IRIS bomb. This scenario follows that of Figure~\ref{fig.georgoulis}, but has additional features to account for the IRIS signatures, as indicated on the figure. From \citet{2014Sci...346C.315P}. Reprinted with permission from AAAS.}\label{fig.hardi}
\end{minipage}
\end{figure}

The line-of-sight (LOS) magnetograms from HMI are valuable for studying the magnetic
field evolution of UV bursts, but the precise magnetic field topology
requires vector magnetograms and/or extrapolation of the field into
the corona. 
Vector magnetograms are available from HMI at 720~second cadence, but
the noise level for the transverse field is about 100~G compared to
about 10~G for the LOS magnetograms, not sufficient for the
small field concentrations that typically underly or cause UV bursts.

A particular scenario believed important for EBs and UV bursts is
U-loop reconnection, in which reconnection takes place in the arms of
a U-shaped magnetic field line (Figures~\ref{fig.georgoulis} and \ref{fig.hardi}). 
Such a field configuration can occur where flux emerges through the
photosphere and mass-loading at the center of the loop prevents it
from rising into the atmosphere. 
When viewed in a LOS magnetogram, the field shows two distant footpoints
and between them an apparent, close bipole where the field is U-shaped
rather than having the $\Omega$ shape of normal field configurations. 
Such configurations are sometimes referred to as ``bald patches''---see the magnetic field lines at the photosphere in Figure~\ref{fig.georgoulis}. 
The difference with $\Omega$ sea-serpent configurations can only be
recognised from vector magnetograms of sufficient quality to
establish that the field connects the two polarities underneath rather than
above. 
This was done by \citet{2002ApJ...575..506G} for an emerging flux
region observed in the Flare Genesis balloon flight, leading them to suggest
the U-shaped loop reconnection scenario for EBs (Figure~\ref{fig.georgoulis}) that was later adopted
also by \citet{2014Sci...346C.315P} for IBs (Figure~\ref{fig.hardi})
Due to this nature of the field configuration the reconnection
necessarily occurs in the low atmosphere.

A different scenario was presented by \citet{2017A&A...605A..49C}, who were able to identify a fan-spine magnetic topology for a MMF UV burst by applying a magnetofrictional relaxation technique to HMI LOS magnetograms. This technique begins with a potential field extrapolation from an initial magnetogram, which is then evolved to a series of non-linear force-free field states by inputting the subsequent magnetograms as new boundary conditions.  The null point of the fan-spine system was found to be at 500~km, and the reconnection driving the UV burst was suggested to be  due to
shearing at the null point, driven by the motion of the MMF.

\citet{2017ApJ...836...52Z} and \citet{2018ApJ...854..174T} combined vector magnetogram data from HMI with a MHD relaxation technique
\citep{2013ApJ...768..119Z,2016ApJ...826...51Z} to
investigate the magnetic environment of UV bursts. 
\citet{2017ApJ...836...52Z} distinguished UV bursts associated with
bald patches and those with flux cancellation. 
\citet{2018ApJ...854..174T} also found that some UV bursts were associated with bald patches for a different emerging flux region, but most of the bursts were not.
A better correspondence was found with 
locations with a high squashing factor 
that are often associated with currents and magnetic reconnection with
regard to the generation of flares.

The evidence from the magnetic field extrapolations is that
reconnection takes place at heights of 0.5--1.0~Mm
\citep{2017A&A...605A..49C,2018ApJ...854..174T}. 
Further evidence for these low heights comes from the weakness or
absence of coronal emission (Section~\ref{sect.cor}) and the
appearance of the narrow cool-gas blends on the IB profiles.
Note that these blends do not require the presence of cool gas on top of
(and part of) the bomb as suggested by
\citet{2014Sci...346C.315P}, 
but may originate in cool gas along a slanted LOS to the
bomb base as in Figure~2 of \citet[][]{2016A&A...590A.124R}. 
Such gas then also absorbs any coronal EUV radiation from the burst at
wavelengths below the hydrogen photoionization edge at 912~\AA.

High-quality vector magnetic field measurements are highly desirable
for UV bursts, both high spatial resolution and high sensitivity. 
Although current ground-based instruments are capable of much better
measurements than HMI, the polarization data are much more susceptible
to bad seeing so that homogeneous good-quality sequences covering the
entire duration of UV bursts have not yet been obtained. 
This is one area where the 4~meter Daniel K.~Inouye Solar Telescope (DKIST)
should offer large improvement. 

Unipolar magnetic field regions show a different type of intensity enhancement, with
examples presented by \citet{2017ApJ...836...63T} for a small EFR. 
In addition to the bursts in the center of the EFR, which were
interpreted as bald patch events, a set of brightenings above the unipolar
patches at the edge of the EFR were identified. 
These were interpreted as shock-heating events driven by plasma
flowing down the legs of the rising arch filament system. 
The unipolar region events showed weaker \ion{Si}{iv} intensities than
the bald patch bursts, with an enhancement of only around five
compared to quiet regions; these may not generally qualify as UV
bursts.


\subsection{Connection to Ellerman bombs}\label{sect.eb2}

\citet{2014Sci...346C.315P} recognized the connection between IBs and EBs by
naming the events ``bombs". 
Their U-loop reconnection scenario was partly based on the similar one
for EBs by \citet{2002ApJ...575..506G}. 
However, without access to simultaneous H$\alpha$ observations, the
authors could not make the IB--EB connection. 
\citep[The later work of ][on the same data-set could, however, make a
connection to EBs for IBs 3 and 4 based on signatures in the
\ion{Mg}{ii} lines---see below.]{2016A&A...593A..32G}

Direct IB--EB connections were demonstrated by
\citet{2015ApJ...812...11V}, \citet{2015ApJ...810...38K} and
\citet{2016ApJ...824...96T}, who each had access to simultaneous
ground-based H$\alpha$ data and found examples of EBs that had a clear
IB signature, i.e., intense, strongly-broadened \ion{Si}{iv} emission
lines with superimposed narrow cool-gas blends.
\citet{2016ApJ...824...96T} identified 10 IBs based on the presence of
broadened \ion{Si}{iv} lines with such superimposed absorption
features. 
Three IBs were clearly matched with EBs, four had no EB signature,
the remaining three were ambiguous. 
They also noted that while the IRIS slit crossed about 30 EBs only 6 had an
IB signature.

\citet{2017A&A...598A..33L} used H$\beta$ to identify 21 EBs, and found four of the events showed wing enhancements in the \ion{He}{i} D$_3$ and \lam10830 lines---the first reported EB signatures in these lines---suggesting temperatures in excess of 20~kK. Two of the four EBs were observed through the IRIS slit and enhanced \ion{Si}{iv} emission was found for each, with one showing IB-like line profiles. This suggests that EBs displaying hot emission may also have signatures in the \ion{He}{i} lines, which could be significant for future ground-based observations (e.g., DKIST).

As simultaneous, high-resolution, spectroscopic H$\alpha$ data are
rarely available for IRIS observations some authors have sought
to find EB identifiers in the IRIS data themselves. 
\citet{2016A&A...593A..32G} found that 1D radiative transfer
models that fit the  H$\alpha$ wings of EBs may also fit the wings of the
\ion{Mg}{ii} lines, and therefore used \ion{Mg}{ii}~h line wing emission to find EBs in their IRIS data-set.
They found that samples  at $-3.5$
and $+1.0$~\AA\ ($-374$ and $+107$~\kms) are suited for identifying
EBs, and subsequently identified 74 events that were a factor two
brighter than their surroundings. 
Around 10\%\ of these events had a signature in the SJI 1400~\AA\ channel.

\citet{2016ApJ...824...96T} used narrow-band filtergrams in the wings
of H$\alpha$ to identify EBs and found
that averaging images obtained at $\pm 1.33$~\AA\ ($\pm 143$~\kms)
from the line core of \ion{Mg}{ii}~k is a good proxy for H$\alpha$
EBs. 
Another method for detecting EBs with \ion{Mg}{ii} was suggested by
\citet{2017ApJ...838..101H} who integrated a region $\pm 0.25$~\AA\
around the two members of the \ion{Mg}{ii} triplet lines at 2798.75
and 2798.82~\AA, and found a good correlation with H$\alpha$ wing
enhancements found with the Fast Imaging Solar Spectrograph on the Goode Solar Telescope.

The ubiquitous availability of AIA full-disk images offers another
avenue for finding EB signatures, with the AIA 1700 and 1600~\AA\
passbands the most promising.  
\citet{2013JPhCS.440a2007R}, \citet{vissers13} and \citet{2015ApJ...812...11V}
compared H$\alpha$ observations from SST/CRISP with AIA and found
good correlations between \Halpha\ EBs and AIA mid-UV brightenings. 
The last paper set as EB criterion $\ge 8\sigma$ above the mean
1700~\AA\ intensity over the whole active region. 
Further study \citep{vissers18} suggests that 1700~\AA\ is the best of the AIA passbands for identifying EBs,
allowing recovery of nearly 20\%\ of the H$\alpha$ EBs when using a $\ge 5\sigma$ above mean intensity threshold, combined with a lower lifetime threshold of 1~minute and size limits of 1--16 pixels. Optimizing instead for the number of AIA candidates that are indeed H$\alpha$ EBs (reaching over 60\%, though only recovering about 5\% of the total H$\alpha$ EB population) requires a higher brightness threshold ($\ge 9 \sigma$) and stricter size constraints (1--9 pixels), but keeping the same lifetime threshold. While complete one-to-one correspondence is thus not possible, the 1700~\AA\  brightenings are of interest in their own right as already noted by \citet{2013JPhCS.440a2007R}.

Finally we emphasize that the results of \citet{2016A&A...593A..32G} and
\citet{2016ApJ...824...96T} suggest that 10--20\%\ of EBs have
an IB signature. 
\citet{2016ApJ...824...96T} have also suggested that 30--60\%\ of UV bursts
are co-spatial with EBs. 
Since Section~\ref{sect.models} below shows that it is
hard to achieve both EB signatures and 
UV burst signatures from a single reconnection event when modeling EBs and IBs, then  obtaining
improved statistics on simultaneous EBs and bursts is desirable.

\subsection{Mini flare ribbons}\label{sect.mini}

\begin{figure}[t]
\centerline{\epsfxsize=\textwidth\epsfbox{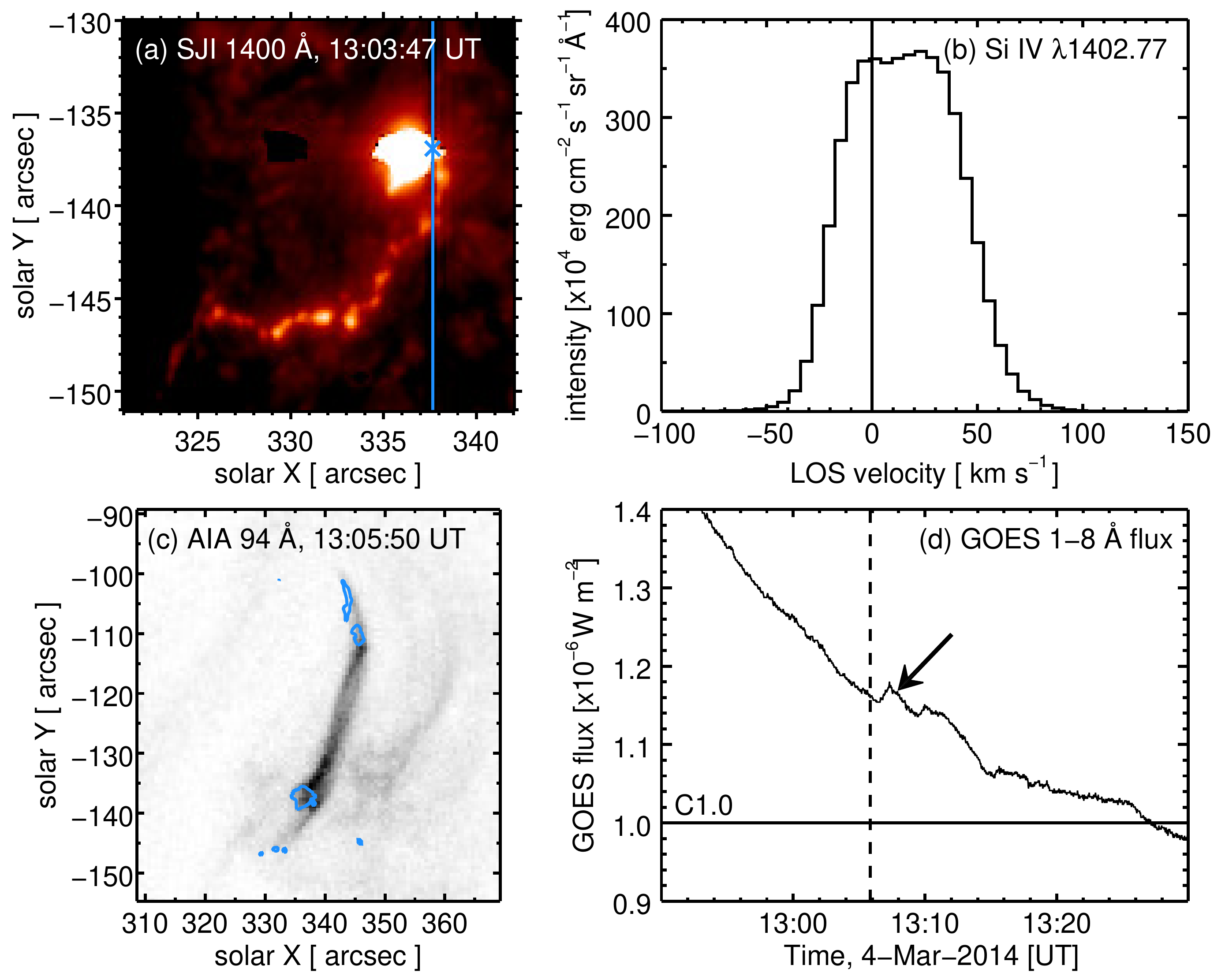}}
\caption[]{\label{fig.flare}
An example of a mini flare ribbon from 2014 March 4. Panel (a) shows an IRIS SJI 1400~\AA\ image with a logarithmic intensity scale. The blue vertical line shows the position of the IRIS slit, and the cross the location corresponding to the spectrum shown in panel (b), where \ion{Si}{iv} \lam1402.77 is shown. Panel (c) shows an AIA 94~\AA\ image (linear intensity scale) from two minutes later, with contours showing the SJI 1400~\AA\ intensity from panel (a) at a level of 1000~DN~s$^{-1}$. Note the displayed region is larger than for panel (a). Panel (d) shows the GOES 1--8~\AA\ flux with the time of the AIA image indicated with the vertical dashed line. An arrow indicates a feature that may correspond to the mini flare (see main text).
}
\end{figure}

Our definition of UV bursts in Sect.~\ref{sect.def} excludes 
bright kernels in flare ribbons, but active
regions can also display what we call ``mini flare ribbons'' that are
small in size (say, 10\as\ or less) and have either no or very weak signal in the 1--8~\AA\ channel of the Geostationary Operational Environmental Satellite (GOES) X-ray monitor.
The \ion{Si}{iv} intensities of such events are comparable to UV
bursts, but a key distinguishing feature is that the lines are mostly
red-shifted, consistent with the profiles seen in most normal flare
ribbons. The ribbons probably correspond to coronal nano- or micro-flares that would be visible at X-ray or EUV wavelengths.

The only example in the IRIS literature is the event described by \citet{2016ApJ...823...60B} which took place in the penumbra of a small sunspot.  Multiple transient brightenings took place along a ribbon of length 10\as, with the brightest comparable with the \citet{2014Sci...346C.315P} IBs.  The \ion{Si}{iv} profiles were mostly Gaussian shaped and red-shifted by 15--20~\kms.  The ribbon also exhibited coronal emission in the AIA channels, which is not typical of UV bursts (Sect.~\ref{sect.cor}).

Similar events have been found by the ISSI team and an example is shown in Figure~\ref{fig.flare}. The IRIS slit crossed a very intense, compact brightening  that is visible in panel (a) and could be interpreted as a UV burst. Note the ``chain'' of weaker brightenings extending from the main brightening that is reminiscent of a flare ribbon. The \ion{Si}{iv} \lam1402.77 line profile is  predominantly red-shifted and significantly narrower than the profiles of Figure~\ref{fig.bursts}. The line's amplitude, however, is comparable to those of Figures~\ref{fig.bursts} and \ref{fig.narrow}. About two minutes after the IRIS image was taken, the AIA 94~\AA\ bandpass (panel c) shows a loop running northwards from the location of the intense IRIS brightening to another strong brightening (not shown in panel a). This loop is clearly filled with plasma at $\sim$~10~MK due to chromospheric evaporation, as expected from the standard flare model. Panel (d) shows the GOES X-ray flux at this time, where a small peak is seen shortly after the time of the AIA image. It is not clear if this peak arises from the event as the GOES flux is an average over the entire solar disk, but if it is then  the energy of an A-class flare is implied.
This event has been studied in more detail  in a recent paper of \citet{2018ApJ...857..137G}.

To distinguish a UV burst from a mini flare ribbon, the key features to check are (i) an extended ribbon structure, (ii) a predominantly red-shifted line profile, and (iii) loop emission in the ``hot'' AIA channels at 94 and 131~\AA\ that would imply chromospheric evaporation has taken place.

\citet{2016ApJ...823...60B} referred to their event as a nanoflare
ribbon after estimating the thermal energy from AIA imaging.
We prefer the term ``mini flare ribbon'' in order to cover a wider
range of energies, and suggest that there is a need for a survey of
such events to investigate how they compare with larger flares.

\subsection{Dynamic loops (arch filament systems)}\label{sect.loops}

\begin{figure}[t]
\centerline{\epsfxsize=\textwidth\epsfbox{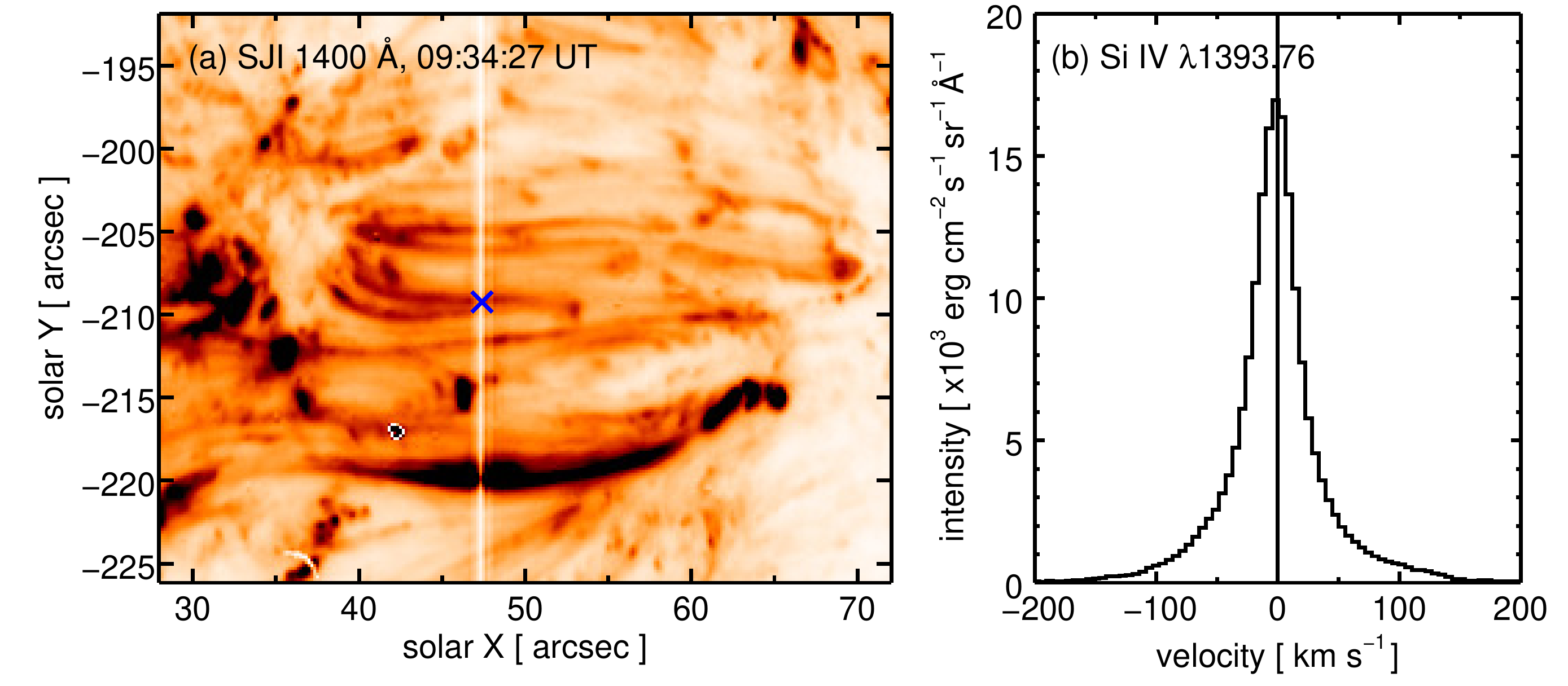}}
\caption[]{\label{fig.loop}
Example \ion{Si}{iv} profile from a dynamic TR loop. The left panel shows an IRIS SJI 1400~\AA\ image of AR 11916 from 2013 December 6. A linear, inverted intensity scale is used. A blue cross denotes the spatial location from which the spectrum in the right panel was taken.
}
\end{figure}

Another class of events in SJI 1400~\AA\ images that show strong UV-line emission but not
considered by us to be UV bursts are dynamic TR loops. These are mostly seen in emerging flux regions, and seem to be analogous to the fibrils of arch filament systems that are seen in absorption in H$\alpha$ \citep{1967SoPh....2..451B}.
Examples have been presented by \citet{2015ApJ...811...48Y}, \citet{2015ApJ...810...46H} and
\citet{2017MNRAS.464.1753H}.
The loops span the emerging flux region and are typically 5--40\as\ long. Propagating intensity fronts in SJI 1400~\AA\ image sequences suggest flows along these loops are common. 
The line profiles often show a relatively narrow central component
with weaker, but very extended wings. We show an example from the data-set studied by \citet{2015ApJ...811...48Y} in Figure~\ref{fig.loop}. These authors showed in their Figure~3 a more intense loop spectrum with a complex line profile. We consider Figure~\ref{fig.loop} to be a more representative dynamic loop spectrum, and Figure~5 from \citet{2017MNRAS.464.1753H} shows a similar example from a different data-set.
Post-flare loop arcades can also exhibit similar profiles, such as
those shown in Figure~4 of \citet{2016ApJ...833..101B}. 
The \ion{Si}{iv} intensities of dynamic loops are generally lower than
those of the most intense UV bursts. For example the peak amplitude of \ion{Si}{iv} \lam1402.77 for the event shown in Figure~\ref{fig.loop} is 9000~\ecss~\AA$^{-1}$, which can be compared with the numbers shown in blue in Figures~\ref{fig.bursts} and \ref{fig.narrow}.
\citet{2017ApJ...836...52Z} applied magnetic field extrapolation to
an emerging flux region (see also Section~\ref{sect.mag}), attributed
the low-lying field lines in their extrapolation to an arcade of dynamic
loops seen in SJI 1400~\AA, and associated the latter with reconnection between the emerging flux
and the overlying magnetic field at quasi-separatrix layers (QSLs). Previously, \citet{2002ApJ...575..506G} had identified the connection between QSLs and EBs for an emerging flux region, and they also identified bright fibrils in the TRACE 1600~\AA\ channel that they interpreted as due to bright \ion{C}{iv} emission. These were likely the equivalent of the \ion{Si}{iv} dynamic loops.

\subsection{Flaring fibrils}\label{sect.FAF}

A study of EBs identified from the Flare Genesis Experiment revealed co-spatial brightenings and dynamic loops in the TRACE 1600~\AA\ channel \citep{2002ApJ...575..506G,2004ApJ...601..530S}. The loops were referred to as ``flaring arch filaments", and \citet{2015ApJ...812...11V} used this notation to describe compact brightenings in AIA 1600~\AA\ images, similar to EBs but with obvious elongated morphology and proper motion
along filamentary strands that they interpreted as due to \ion{C}{iv}
line emission lines in this passband. \citet{2016A&A...590A.124R} 
subsequently proposed that ``flaring active-region fibrils'' (FAFs) is a
better name as it avoids confusion with the dynamic loops of arch filament systems referred to in the previous section.

\begin{figure}[t]
\centerline{\epsfxsize=\textwidth\epsfbox{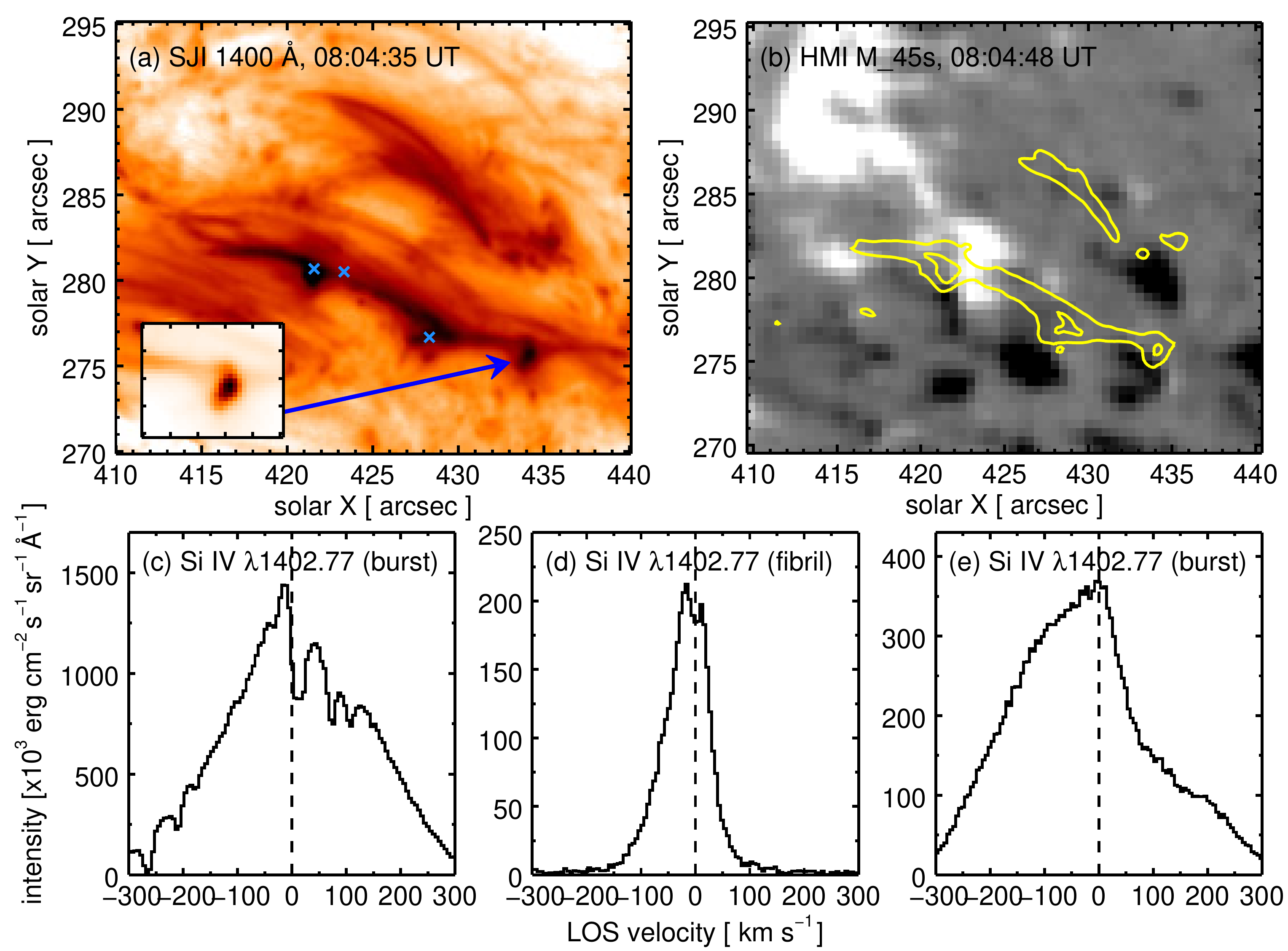}}
\caption[]{
Example of a flaring arch fibril. Panel (a) shows an IRIS SJI 1400~\AA\ image of AR 12089 from 2014 June 15. A logarithmic, inverted intensity scale is used, and the inset shows a section of the image containing a FAF with a linear intensity scale. Panel (b) shows a co-temporal LOS magnetogram from HMI, scaled between $-500$ and $+500$~G. Yellow contours are derived from the SJI image, with levels of 300 and 2000~DN~s$^{-1}$. Panels (c), (d) and (e) show three \ion{Si}{iv} \lam1402.77 spectra corresponding to the left, middle and right locations indicatd by the blue crosses in panel (a).
}\label{fig.faf}
\end{figure}

\citet{2015ApJ...812...11V} compared AIA FAFs with IRIS data and found
intense, IB-like \ion{Si}{iv} profiles at the locations of the
1600~\AA\ brightenings, with jets or loops in the SJI 1400~\AA\
images and shell-like fronts expanding away from them in the 
hotter AIA passbands.
To highlight the difference between the dynamic loops described in the previous section with FAFs, we show a SJI 1400~\AA\  image from the 2014 June 15 dataset studied by 
\citet{2015ApJ...812...11V} in Figure~\ref{fig.faf}a. There are many dynamic loops in this image, and they can be seen to have a generally smooth variation in their intensities from footpoint to footpoint. The FAFs are the brightest loops in the image which appear to be connected to compact brightenings. Note that the brightenings can be identified as discrete features---the inset image of Figure~\ref{fig.faf}(a) shows one of the FAF brightenings on a linear intensity scale, revealing a distinct brightening. In particular, the brightenings are not simply small segments  that brighten during the loop's evolution. \ion{Si}{iv} \lam1402.77 spectra from three locations are shown in panels (c)--(e). (c) and (e) correspond to FAF brightenings, and show the broad, very intense profiles of IBs (compare with Figure~\ref{fig.bursts}), although (e) does not show the cool absorption blends. Panel (d) shows a profile from a fibril that connects to the FAF brightening, and it is more comparable to the dynamic loop profile shown in Figure~\ref{fig.loop}, although the intensity is an order of magnitude larger. The FAF brightenings we consider to be UV bursts, and a plausible interpretation is a low-lying reconnection event that is able to connect to the overlying arch filament system and deposit heated plasma in certain loops.

Another event in the literature that we consider to be a FAF is the one shown in Figure~13 of \citet{2015ApJ...810...46H}, which was described as the footpoint of two interacting loop systems that exhibited explosive event line profiles.

To summarize the terminology, the FAF is the combination of the loop (fibril) and the burst that evolve together.

\subsection{Coronal signatures}\label{sect.cor}

\citet{2014Sci...346C.315P} noted that the four IBs they observed did not have a signature
in the coronal AIA passbands.
This seems to be a common feature of UV bursts, but not an absolute rule. 
For example, the event of \citet{2015ApJ...809...82G} showed weak AIA
signals, allowing the authors to derive a differential emission
measure curve, and 
\citet{2015ApJ...812...11V} 
described thin long arcs seen in the hotter AIA diagnostics spreading
away from FAF sites.

The weakness or absence of co-spatial coronal emission either implies
that the bursts are not heated beyond about $10^5$~K, or that the EUV
emission is blocked by overlying cool plasma with sufficient neutral
hydrogen and helium that all lines below 912~\AA\ are strongly
attenuated. IRIS does observe a coronal line above 912~\AA---\ion{Fe}{xii} \lam1349.40, with $T_{\rm max}=1.6$~MK---but this line is very weak and has not yet been reported from a UV burst observation.

\section{UV burst modeling}\label{sect.models}

Our interpretation of UV bursts as magnetic reconnection events in the low atmosphere means there are three components to modeling that must be considered: (1) the atmospheric evolution that leads to reconnection at low heights in the atmosphere; (2) the physics associated with the reconnection current sheets; and (3) the effect of dynamics and heating on spectral line profiles.
Our definition of a UV burst requires the simulations to produce a compact, very intense, flickering brightening in a TR line (with $T_{\rm max}\sim 100$~kK). Since most UV bursts also show the complex line profiles of IBs and many also the cool-line absorption blends, then these are additional tests for the models to pass. 

UV bursts are relatively new discoveries so simulations directly focused on them are few, but a number of works have considered how simulations relevant to EBs may produce UV bursts. In particular, over the past 20 years increasingly sophisticated codes for modeling flux emergence have been performed and demonstrated to produce the U-loop reconnection that has been suggested for EBs and IBs (Figures~\ref{fig.georgoulis}, \ref{fig.hardi}). The physics of magnetic reconnection current sheets has also been studied extensively, and may be particularly relevant to the often complex line profiles of UV bursts. Another focus of modeling has been on whether 1D radiative transfer models that are able to reproduce the profiles of strong chromospheric lines from EBs are consistent with UV bursts.

1D modeling of EBs has a history going back to \citet{1983SoPh...87..135K}, with the most basic aim to reproduce the wings-only brightening of H$\alpha$. The method 
involves the perturbation of a plane-parallel atmosphere, usually by 
the insertion of a hot component
near the temperature minimum region (heights of $\approx 450$~km). 
NLTE radiative transfer is included and profiles of strong lines
such as H$\alpha$, \ion{Ca}{ii} \HK\ and \ion{Mg}{ii} \hk\ are
modeled. 
Recent work in this vein includes 
\citet{2010MmSAI..81..646B}, 
\citet{2014A&A...567A.110B}, 
\citet{2016A&A...593A..32G}, 
\citet{2017RAA....17...31F} and \citet{vissers18}, 
and the more sophisticated 2D NLTE modeling by
\citet{2013A&A...557A.102B}. 
Cloud modeling introduces an additional plasma component to simulate the effect of absorption by overlying chromospheric fibrils on the line core, and a recent example is the work of
\citet{2017ApJ...838..101H}. 
Comparisons of the models with observations typically constrain the hot component to be about 100--3000~K above the background temperature \citep[e.g.,][]{2015RAA....15.1513L,2016A&A...593A..32G}, although \citet{vissers18} found some cases of localized enhancements of 10--15~kK from EB/UV burst inversions. Note that, since the models are focused on effects at the temperature minimum region, the atmospheres are usually truncated at 20\,000~K or lower, well below the $T_{\rm max}$ of \ion{Si}{iv}.

The discovery of IBs led \citet{2017RAA....17...31F},
\citet{2017ApJ...835L..37R} and \citet{2017ApJ...845..144H} to
investigate whether the EB models could be modified to produce
\ion{Si}{iv} emission. \citet{2017RAA....17...31F} considered
temperatures of the hot component up to 15~kK, but then the
chromospheric signatures were not consistent with the EB
observations. Both \citet{2017ApJ...835L..37R} and
\citet{2017ApJ...845..144H} used the RADYN code
\citep{1997ApJ...481..500C}, which couples radiative transfer with
hydrodynamics, to allow the plasma to respond to the heating. Again
the authors found that it was not possible to reconcile \ion{Si}{iv}
emission with the chromospheric line profiles of EBs. 
On the other hand, recent EB/UV burst inversions by \citet{vissers18} have achieved agreement between \ion{Ca}{ii} \lam8542, \ion{Ca}{ii} K and \ion{Si}{iv} (but not simultaneously with \ion{Mg}{ii} h \&\ k) with temperature profiles peaking at 10~kK in the low atmosphere and/or enhancements to 35--40~kK at the base of the chromosphere. Further synthesis tests also suggested that a temperature enhancement of 30-50 kK at low chromospheric heights could be sufficient to get the \ion{Si}{iv} emission right, with minimal effects on the goodness-of-fit of the \ion{Ca}{ii} \lam8542, \ion{Ca}{ii} K  and \ion{Mg}{ii} h \&\ k lines.
We highlight
that, as discussed in Sect.~\ref{sect.eb2}, only 10--20\%\ of EBs have
been found to have a hot UV burst signature, thus the failure of the
1D models to account for the hot emission does not invalidate the
models for the much more common, cooler events. In addition, it is
possible that non-equilibrium effects such as $\kappa$ electron
distributions can cause \ion{Si}{iv} to be formed at much lower
temperatures \citep{2014ApJ...780L..12D}. A similar effect may also
arise if EBs are formed at sufficiently high density that \ion{Si}{iv}
has near-Saha-Boltzmann opacity, 
giving a formation temperature of
10--20~kK \citep{2016A&A...590A.124R}.
However, within this cooler regime no attempt has been yet made to
reproduce the non-visibility of EBs in the \ion{Na}{i}\,D and
\ion{Mg}{i}\,b lines, already mentioned by
\citet{1917ApJ....46..298E} 
and confirmed by \citet{2015ApJ...808..133R}, 
which may be difficult to reconcile with 1D temperature humps covering
the formation heights of these lines.

\begin{figure}[t]
\centerline{\includegraphics[width=0.8\textwidth]{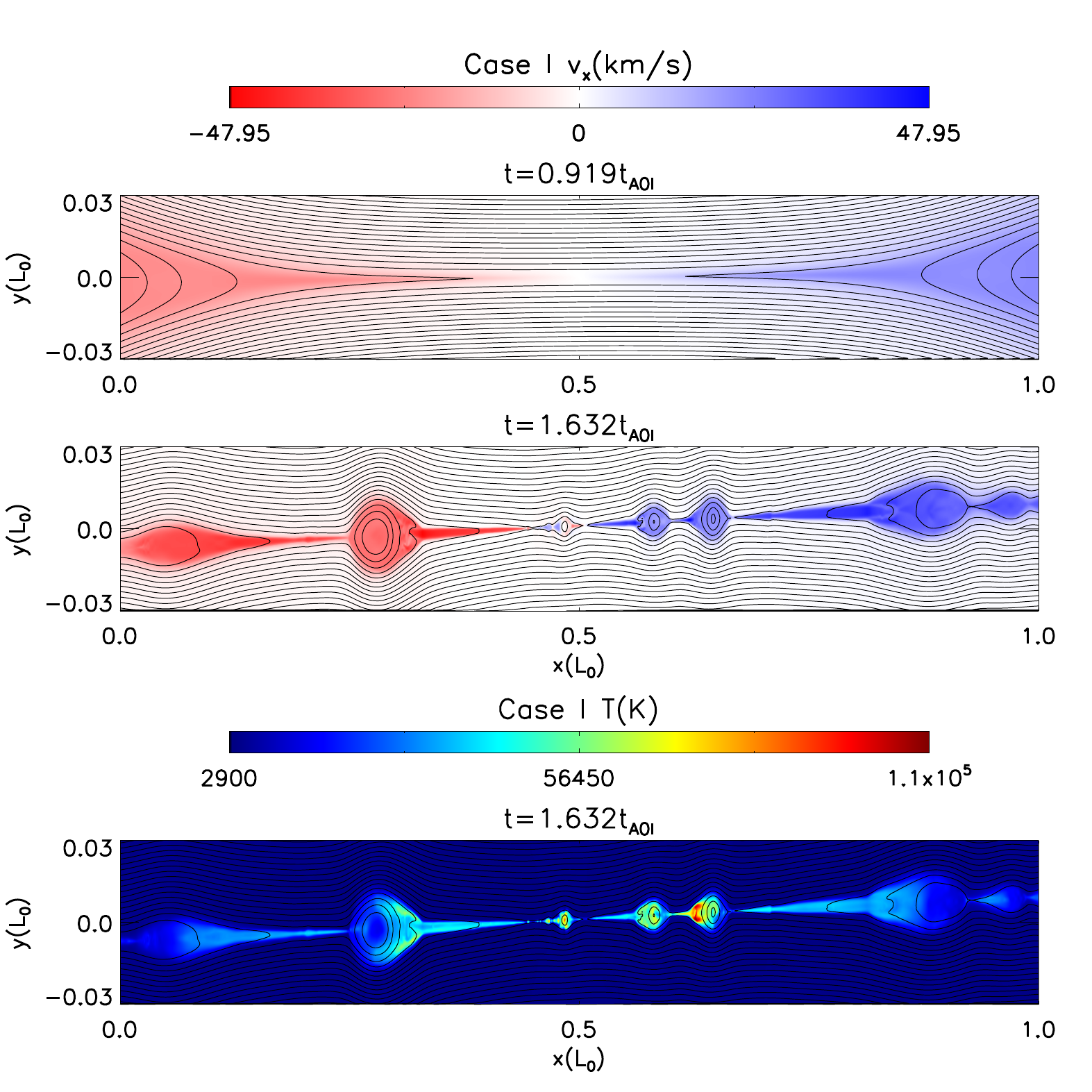}}
\caption[]{Images from the current sheet simulation of \citet{2016ApJ...832..195N}. The upper panel shows the configuration prior to plasmoid formation, with velocity in the $X$-direction ($v_x$, parallel to the current sheet) plotted; contours show the magnetic field. The middle panel shows $v_x$ at a later time after plasmoids have formed, and the lower panel shows temperature at this time, revealing 100~kK temperatures in the plasmoids.
Reproduced by permission of the AAS.}
\label{fig.leini}
\end{figure}

There are a number of simulations that focus solely on the physics of magnetic reconnection current sheets in the solar atmosphere. They do not investiate the wider plasma evolution  that leads to the sheet formation, but they do resolve the current sheet at  higher resolution than large-scale MHD codes.
\citet{2015ApJ...799...79N,2016ApJ...832..195N} investigated plasma
heating in current sheets located at atmospheric heights of 100 to 1000~km, and with horizontal and vertical orientations. Plasmoids are formed in the current sheet, and heating occurs where slow-mode shocks from the reconnection site interact with the plasmoids (Figure~\ref{fig.leini}). They found that the  plasma $\beta$  (ratio of plasma pressure to magnetic
pressure) is crucial in determining if UV burst temperatures are
produced: $\beta \lesssim 0.1$ enables heating to 80~kK (the temperature of formation of \ion{Si}{iv}). The further work of \citet{2018ApJ...852...95N} found that the inclusion of non-equilibrium ionization in current sheet simulations results in a faster reconnection rate but smaller temperature increases, potentially affecting the formation of \ion{Si}{iv}.

\citet{2015ApJ...813...86I} also studied current sheet dynamics in relation to burst-like events recorded from IRIS. They highlighted that the break-up of a current sheet into plasmoids separated by magnetic field x-points at which particles are accelerated can give rise to complex velocity distributions. Modeling the \ion{Si}{iv} emission lines integrated along the current sheet was found to give realistic, IB-like profiles. They also highlighted that a relatively small change to the line-of-sight had a significant effect on the profiles.

\begin{figure}[t]
\includegraphics[width=\textwidth]{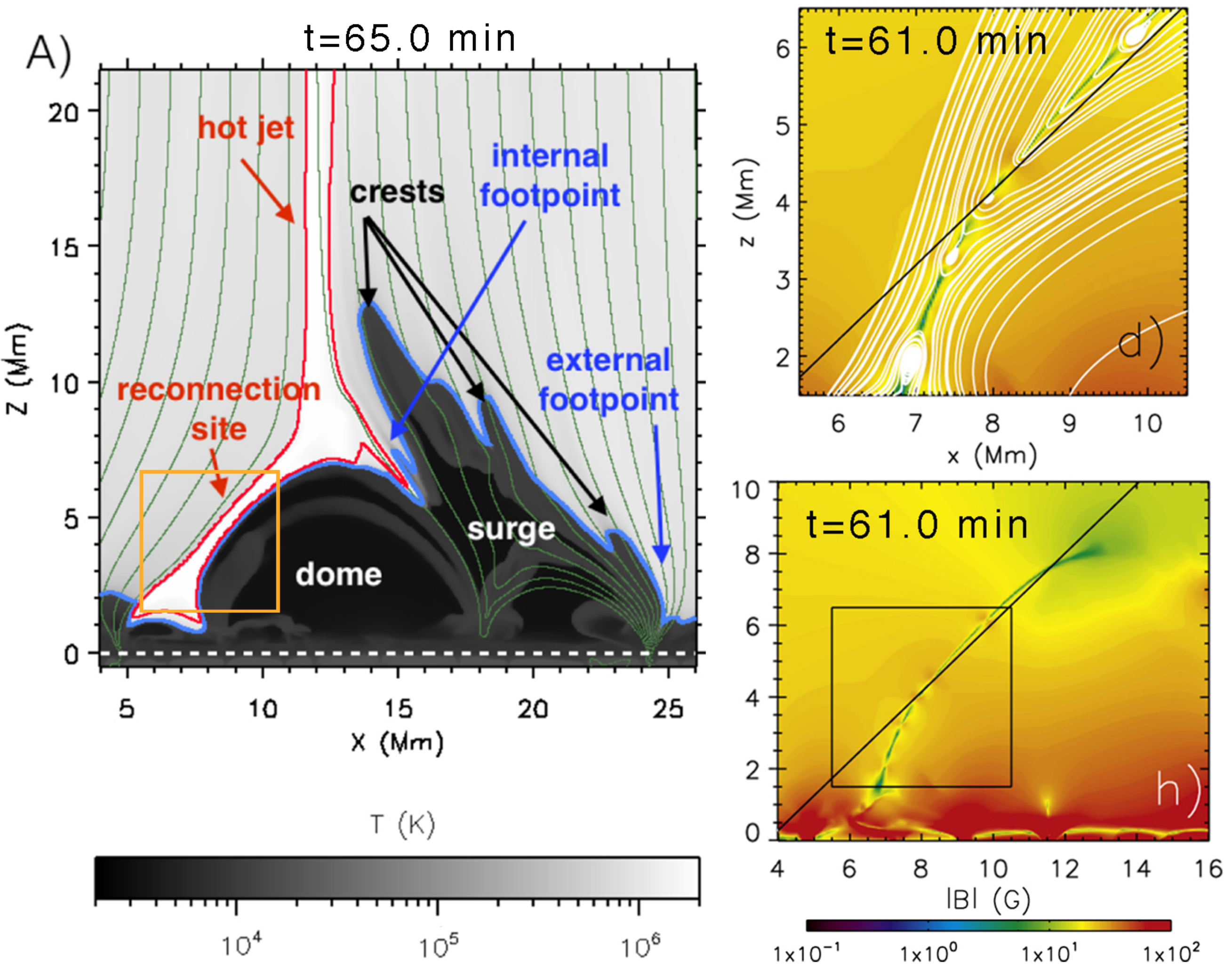}
\caption[]{Still frames from a 2.5D simulation of emerging flux. The left panel \citep[from][]{2017ApJ...850..153N} shows a temperature map with green contours representing magnetic field, the blue contour shows the layer at which \ion{Si}{iv} is formed (80~kK), and the red contour shows plasma at 1.2~MK. Key features of the model are labeled. The orange box highlights the location of the current sheet, which is shown in progressively greater detail in the bottom-right and top-right panels, which both show the magnetic field strength. Magnetic field contours on the top-right panel emphasize the location of the plasmoids along the current sheet. These panels are from \citet{2017ApJ...851L...6R} at a slightly earlier time than the left panel.
Reproduced by permission of the AAS.}
\label{fig.2dmodel}
\end{figure}

Another 2.5D simulation that produced current sheet plasmoids, but this time as part of the wider evolution of solar atmospheric structures was presented by \citet{2017ApJ...850..153N} and \citet{2017ApJ...851L...6R}. The interaction of an emerged bipole with pre-existing magnetic field was simulated, leading to a surge  and a current sheet. The latter occurred where the ``dome'' of the emerged flux pushed against opposite polarity field, and the simulation resolution was sufficient to resolve plasmoids. Figure~\ref{fig.2dmodel} shows images taken from these papers. The authors chose the line-of-sight indicated by the diagonal lines in the two right panels, and they found that complex, IB-like \ion{Si}{iv} profiles were produced from the multiple plasmoids along the current sheet.
Note that the reconnection in this scenario is not the U-loop reconnection discussed in Sect.~\ref{sect.mag}, but that more commonly associated with jets \citep[e.g., Figure~8 of ][]{2000ApJ...542.1100S}. This is also the scenario inferred by \citet{2017A&A...605A..49C} for a MMF burst.

There are many time-dependent simulations of emerging flux in the literature \citep[see review of][]{2015sac..book..227S}, and a particular issue relevant to UV bursts and EBs is the occurrence of U-loop reconnection (Sect.~\ref{sect.mag} and Figures~\ref{fig.georgoulis}, \ref{fig.hardi}). This has been demonstrated to arise naturally on account of the Parker buoyancy instability acting on the flux and giving rise to so-called serpentine field lines.
A 2D resistive MHD simulation of an emerging flux sheet was presented by \citet{2007ApJ...657L..53I}, and it showed plasma heating and flows at the sites of U-loop reconnection that were identified with EBs.
A similar result was found by \citet{2009A&A...508.1469A} but for a fully-3D resistive MHD simulation of flux sheet emergence.
Further advances were made by \citet{2009A&A...507..949T} and \citet{2010ApJ...720..233C}, who performed 3D simulations for the emergence of twisted cylindrical and semi-torus flux tubes, respectively, demonstrating U-loop reconnection in both cases. The latter work effectively demonstrated that EBs can occur during the formation of a bipolar active region.

These latter two simulations both used the MURaM code \citep{2005A&A...429..335V} and included simplified radiative transfer effects. More recently the code has been applied to the case of an emerging flux sheet by \citet{2017A&A...601A.122D}, but with a more sophisticated treatment of radiative transfer, including detailed H$\alpha$ synthesis. The model produces realistic EB signatures, but the atmosphere does not extend beyond the chromosphere and so UV bursts are not produced.

The Bifrost code of
\citet{2011A&A...531A.154G} does extend to the chromosphere and corona, allowing the evolution of flux emergence to be studied in these layers. 
This was done in full 3D by
\citet{2017ApJ...839...22H} 
who explicitly address EBs, UV bursts including IBs, and microflares. 
It includes spectral synthesis of \Halpha, \ion{Mg}{ii}\,\hk, and the
\ion{Si}{iv}\, 1393.7\,\AA\ line including the \ion{Ni}{ii} blend for
selected simulation snapshots. 
There are many striking agreements between the simulated and observed
appearances of both EBs and UV bursts. 
In the simulation the EBs are photospheric in origin and remain
generally under 10\,000\,K; the IB-like UV bursts are chromospheric and
reach higher temperatures although their total energy-release content
is far less. 
The microflares go higher up and reach coronal temperatures.
In this simulation they are all the result of reconnection between the
legs of dragged-down U-loops and expanding $\Omega$-loops that are part
of serpentine flux emergence.
Unlike the 2D codes discussed earlier, the simulation resolution is not sufficient to resolve plasmoids in the current sheets, but broad \ion{Si}{iv} line profiles are produced from bi-directional flows at the reconnection sites.
A remaining discrepancy is that the simulated EBs show \ion{Si}{iv}
emission only barely or none at all, whereas this has been observed
for 10--20\%\ of EBs. This version of the Bifrost code did not
yet account for ion-neutral separation (ambipolar diffusion) whereas
tests have shown in the meantime that this can substantially increase
the rate of reconnection
\citep{2017ApJ...847...36M}, which may impact the predicted
emissions of EBs. 

\section{Summary}\label{sect.summary}

The IRIS instrument has revealed a set of intense, transient, compact  brightenings that occur in active regions and are visible in the resonance lines of \ion{Si}{iv}.  In this work we advocate that these events be referred to as UV bursts.  The definition of UV bursts is given in Section~\ref{sect.def} and it includes the set of events that have been called IRIS bombs in the literature.  The latter were introduced by \citet{2014Sci...346C.315P}, and are defined through their spectroscopic signature in the \ion{Si}{iv} lines.

Studies of UV bursts are ongoing, but they are widely believed to be
small-scale reconnection events occurring at heights from the
photosphere to the upper chromosphere. 
As the events reach TR temperatures then they have special advantages
for investigating reconnection in the solar atmosphere. 
The TR is thin (typically 10's of kilometers) and so the large energy
output from reconnection gives a very strong signal in TR lines that
overwhelms the background emission and allows fine details in the line
profiles to be studied. 
In contrast a reconnection event in the corona takes place within a
volume that is 10's of megameters thick, and thus gives a relatively
subdued signal unless it is a significant size flare. 
The TR has a number of very strong lines in the far ultraviolet where
instrument sensitivity, spectral and spatial resolution are generally
higher than for the EUV and X-ray regions, where most coronal lines
are found. 
In the chromosphere, the lines are mostly formed optically thick, and thus
reconnection events in this region are mostly studied only in the
lines' wings. 
However, the Fabry-P\'erot spectroscopic imaging instruments commonly
used have relatively narrow wavelength coverage and thus do not cover
the full spectral extents of the lines. 
The TR lines in the UV therefore offer excellent opportunities for
studying the physical processes of magnetic reconnection and heating
in the solar atmosphere. 

The early work on UV bursts has suggested certain common features: they mostly have weak or non-existent signal in the coronal imaging channels of AIA; they generally overlie small-scale, dynamic magnetic features seen in LOS magnetograms from HMI, and emerging flux regions, moving magnetic features and light bridges are the most common locations; and  about 10 to 20\%\ of UV bursts are cospatial with EBs, which are generally interpreted as photospheric reconnection events.

A number of modeling efforts have already been applied to UV bursts with some success (Section~\ref{sect.models}), but one common feature is that it has not been possible to achieve both an EB signature and a UV burst signature from the same event.

Despite the progress already made in describing and understanding UV bursts, we identified some areas for further work with the IRIS data in the preceding text that we summarize here.

\begin{itemize}
  \item Coordinated studies with Hinode/EIS and IRIS to determine the maximum temperatures reached by the events.
  \item IRIS imaging and spectroscopy at the maximum possible cadence of 2~seconds to investigate the timescale of intensity fluctuations.
  \item A wider survey of narrow-line bursts and their connections with the magnetic field.
  \item A statistical study of bursts in the AIA 1600 and 1700~\AA\ bandpasses during the lifetime of an active region.
  \item Continued statistical studies of the EB--UV burst connection using H$\alpha$ data, and the identification of definitive EB signatures in the IRIS chromospheric lines.
  \item A survey of mini flare ribbons and their properties.
\end{itemize}

The next solar UV spectroscopic capability for solar physics is the SPICE instrument \citep{2013SPIE.8862E..0FF} that will be flown on Solar Orbiter.  This will observe two wavelength bands at 704--790 and 973--1049~\AA, which give excellent coverage from the chromosphere through the transition region to the corona, and thus will allow the UV burst heating to be tracked through the layers of the atmosphere.  The spatial and spectral resolutions will be somewhat lower than those of IRIS, but will  be sufficient to see UV bursts.

The Atacama Large Millimeter/submillimeter Array \citep[ALMA;][]{2016SSRv..200....1W} promises superb new capability for UV-burst studies
when the (far from trivial) technological problems of achieving long-baseline, many-telescope observations including full calibration become solved.  Since hydrogen ionizes in UV bursts, their free-free hydrogen opacity becomes very large, even exceeding Lyman-$\alpha$ at high temperature, and since free-free extinction is an LTE process ALMA will directly 
measure temperatures of UV bursts wherever these are optically thick, at outstanding angular resolution.

DKIST is expected to become operational in late-2019/early-2020 and will resolve structures down to 25~km size and enable high quality vector magnetograms through higher polarimetric sensitivities compared to other ground-based telescopes. DKIST will have no transition region capability, but the \ion{He}{i} D$_3$ and \lam10830 lines may be proxies (Sect.~\ref{sect.eb2}), and EBs will be 
prime targets that offer the opportunity to study the reconnection process in unprecedented detail. A very recent study of a UV burst at 75~km resolution \citep{2018arXiv180701078S} with the \ion{Ca}{ii} H line from the S{\sc unrise} observatory suggests bursts may have a complex, flare-like structure when observed at ultra-high resolution.

Longer term, a new UV/EUV spectroscopic mission with improved spatial resolution compared to IRIS and a more complete temperature coverage would be ideal for UV bursts. The rapid variability of UV bursts  makes sub-second temporal cadence desirable, which is currently limited by the readout times of UV-sensitive CCDs. The recent report\footnote{http://hinode.nao.ac.jp/SOLAR-C/SOLAR-C/Documents/NGSPM\_report\_170731.pdf} of the Next Generation Solar Physics Mission Science Objectives Team (NGSPM-SOT) recommended a coronal/transition region spectrograph with 0.3\as\ spatial resolution, and  \citet{2012ExA....34..273T} presented an instrument concept called LEMUR (Large European Module for solar Ultraviolet Research) that is compatible with the NGSPM-SOT recommendations. Such an instrument would provide new advances for UV burst studies, particularly when combined with DKIST.

%
\begin{acknowledgements}
The authors thank ISSI Bern for support for the team ``Solar UV
bursts---a new insight to magnetic reconnection''.
P.R.~Young acknowledges funding from NASA grant NNX15AF48G, and thanks the Max Planck Institute in G\"ottingen for kind hospitality during visits in 2015 and 2016.
H.~Tian is supported by NSFC grant 41574166, the Max Planck Partner
Group program, and the Recruitment Program of Global Experts of
China. 
The work was supported by JSPS KAKENHI Grant Numbers JP16K17671 (PI: S.~Toriumi), JP15H05814 (PI: K.~Ichimoto), and JP25220703 (PI: S.~Tsuneta).
G.~Vissers acknowledges support under the CHROMOBS grant from the Knut and Alice Wallenberg Foundation.
L.~Rouppe van der Voort's research is supported by the Research Council of Norway, project number 250810, and through its Centres of Excellence scheme, project number 262622.
Z.~Huang thanks the support from the NSFC (41404135) and the Young Scholar Program of Shandong University, Weihai (2017WHWLJH07).
C.J.~Nelson is thankful to the Science and Technology Facilities Council for support received.
L.P.~Chitta received funding from the European Union's Horizon 2020 research and innovation programme under the Marie Sk\l{}odowska-Curie grant agreement No.\,707837.
A.~Berlicki and P.~Heinzel acknowledge support by the grant No.~16-18495S of the
Czech Funding Agency and by ASI ASCR project RVO: 67985815.
We thank L.~Kleint for useful discussions.  IRIS is a NASA
small explorer mission developed and operated by LMSAL with mission
operations executed at NASA Ames Research center and major
contributions to downlink communications funded by the Norwegian Space
Center (NSC, Norway) through an ESA PRODEX contract.  
\end{acknowledgements}

%
%
\bibliographystyle{spr-mp-sola}
\bibliography{ms,ms_ads}

%
 \appendix   

\section{Transition region imaging}\label{sect.imaging}

Imaging the solar atmosphere through narrow bandpass filters enables
particular temperature layers to be isolated and studied at high
temporal resolution.  At visible wavelengths, Fabry-P\'erot filters have
long been used and they enable high cadence spectral imaging in
chromospheric lines
such as H$\alpha$ and \ion{Ca}{ii} H \& K.  For shorter wavelengths ($\approx$ 100--500~\AA)
multilayer coatings on optical surfaces enable good reflectivity over
a relatively narrow wavelength range (typically 10\%\ of the central
wavelength).  If the filters are centered on strong, isolated emission
lines then the images can be a good means of isolating a particular
temperature regime in the atmosphere. This has been exploited
by several space-based instruments with the most advanced being  SDO/AIA, which obtains
full-disk solar images in seven EUV filters between 94 and
335~\AA\ with around 1~arcsec spatial resolution and 
a 12~second cadence.
The 100--500~\AA\ region contains mostly coronal emission lines, with
the only strong, isolated cooler lines being \ion{He}{ii} \lam304 and
\ion{Ne}{vii} \lam465. 
The former is observed with AIA and is formed around 80~kK, however
the formation of this line is optically thick, is complex and so the
emission not easy to interpret. 
The \ion{Ne}{vii} line is from the upper transition region (0.5~MK),
but has not been successfully observed with a multilayer instrument
yet.

The lack of filter imaging for the TR   means that imaging has  been restricted to
spectrometers that have the option of a wide slit (usually referred to
as a slot) that allows
monochromatic imaging in a small spatial region but at the risk of
overlapping images from 
neighboring emission lines.  SOHO/CDS had the option of a
90\as\ $\times$ 240\as\ slot that yielded transition region images in
\ion{O}{v} \lam629.7 although only at 6--10\as\ spatial resolution
\citep[e.g.,][]{2000A&A...353.1083B}.  More recently Hinode/EIS has the
options of 40\as\ and 266\as\ slots, although there are no strong
transition region lines.  Useful data have been obtained from the upper
TR lines
\ion{Mg}{vi} \lam269.0 and \ion{Si}{vii} \lam275.4, however
\citep{2009ApJ...695..642U}.  Since using a narrow or a  wide slit is
an either/or option and spectrometer science is typically focused on
emission line parameters, then TR imaging has been rare.  This has had
the consequence that transition region phenomena have generally been classified based on
their spectroscopic signature.  For example, explosive events
(Sect.~\ref{sect.ee}) are principally distinguished by their line profiles. 

The first filtergraph  imager for the transition region is the IRIS
slitjaw imager (SJI), which has a filter centered at 1390~\AA\ that
picks up the \ion{Si}{iv} lines at 1393.8 and 1402.8~\AA, as well as the FUV continuum.
For intense UV bursts the \ion{Si}{iv} emission dominates the signal in the 1400~\AA\ images and so the bursts can be confidently assigned to the transition region.

\section{Emission measure comparison}\label{sect.em}

The large \ion{Si}{iv} intensities of UV bursts are their most striking
feature, and they imply a large emission measure (EM).  Here we compare a typical
UV burst EM with values derived from the earlier features that seem most
similar to them, namely the SMM/UVSP bursts (Sect.~\ref{sect.bursts})
and active region blinkers (Sect.~\ref{sect.blinkers}).

The expression relating an observed intensity, $I$, to a volume emission
measure, ${\rm EM}_V$, for an isothermal plasma of temperature $T$ is
\begin{equation}\label{eq.em}
I = G(T,N_{\rm e})  { {\rm EM}_V \over A}
\end{equation}
where  $G(T,N_{\rm e})$ is
the contribution function for the observed emission line that is
computed using the CHIANTI \verb|gofnt| IDL software routine.  For the
lines considered here, $G$ has a weak dependence on density and we
compute the function at a density of $10^{12}$~cm$^{-3}$ and the
temperature at which the function peaks.  $G$ contains
the element abundance of the emitting ion and we use the ``Caffau''
solar photospheric abundance set from CHIANTI
\citep[see][]{caffau11,chianti71}.  $A$ is the area of the event, and
the interpretation varies depending on the spatial resolution of the
instrument, as described below.  

For IRIS we consider event 1 of  \citet{2014Sci...346C.315P} -- see also panel (b)
of Figure~\ref{fig.bursts}.  For
computing the ${\rm EM}_V$ value we sum the intensity over a
$3\times 7$ block of spatial pixels centered on the peak intensity
value, and we sum across the spectral line profile to obtain the
intensity listed in Table~\ref{tbl.em}.  The intensity profile across
the burst in the $Y$-direction takes a Gaussian shape with a
full-width at half-maximum of 0.62\as.  We thus assume that the event
has a spatial size of 0.62\as\ $\times$ 0.62\as, which we use as $A$
in Eq.~\ref{eq.em}.

\citet{hayes87} give \ion{Si}{iv} \lam1402.7 intensities for nine
bursts in their Table~1, and we use the median value  for Table~\ref{tbl.em}. The
UVSP pixels for this observation had a size 
4\as\ $\times$ 4\as\ and we assume the bursts were fully contained
within these pixels, so $A=$4\as\ $\times$ 4\as.

For SOHO/CDS we choose the event reported by \citet{young04} from the
data-set s7616r00 from 1997 April 15, beginning at 21:12~UT.  This was
a sit-and-stare observation, and the intensity was measured from the
exposure with the peak brightness in the \ion{O}{v} \lam629.7
line.  The burst produced a Gaussian distribution of intensity along
the CDS slit, with a FWHM consistent with the PSF of the instrument
\citep{young04}.  Therefore $A$ was set to the CDS pixel size,
which in this case was 4.00\as\ $\times$ 1.68\as.

\begin{table}
\caption{Emission measure parameters for a selection of intense transition region brightenings.}\label{tbl.em}
\begin{tabular}{ccccccll}     
\hline
  &
  Wavelength &
  Intensity &
  $G$ &
  $A$ &
  EM$_V$ \\
  Ion &
  (\AA) &
  (\ecss) &
  (erg~cm$^3$~s$^{-1}$~sr$^{-1}$) &
  (cm$^2$) &
  (cm$^{-3}$) &
  Instrument &
  Reference\\
  \hline
  \ion{Si}{iv} & \lam1402.8 & 1.86(+6)
  & $1.77(-25)$ & $2.02(+15)$ & $2.12(+46)$ 
  & IRIS & \citet{2014Sci...346C.315P} \\
&&1.80(+5) 
  & $1.77(-25)$ & $8.41(+16)$ & $8.55(+46)$ 
  & SMM/UVSP & \citet{hayes87} \\
\ion{O}{v}  & \lam629.7 & 5.27(+5)
   & $9.56(-24)$ & $3.53(+16)$ & $1.95(+45)$ 
   & SOHO/CDS & \citet{young04} \\
\ion{O}{v}  & \lam192.9 & $8.53(+3)$
   & $3.88(-25)$ & $1.05(+6)$ & $2.31(+44)$ 
   & Hinode/EIS & \citet{young07-tr} \\
\hline
\end{tabular}
\end{table}

The Hinode/EIS event of \citet{young07-tr} was chosen for deriving an
emission measure and, to enable direct comparison with CDS, the
intensity of 
\ion{O}{v} \lam192.9 was measured.  The value of $G$ was calculated
by including the two transitions at 192.904 and 192.911~\AA.  As the
intensity distribution along the EIS slit had a width consistent with
the instrument point spread function, then we considered the event to
be unresolved, so the intensity was summed in the $Y$
direction and $A$ was set to 2\as\ $\times$ 1\as, the pixel size for
this observation.

Comparing the EM values in Table~\ref{tbl.em} we see
that the \ion{Si}{iv} values are at least an order of
magnitude higher than the CDS \ion{O}{v} value.  We caution
that \ion{Si}{iv} belongs to the sodium isoelectronic sequence, and
there is a well-known problem that ions from the sodium and
lithium-like isoelectronic sequences generally yield EM values
higher than other species at similar
temperatures \citep{1972ApJ...178..527D}.  However, this is unlikely to account for an order of
magnitude difference and it suggests a significantly smaller EM in the
upper transition region.  The EIS event is a further order of magnitude
weaker than the CDS event and this, coupled with the apparent lack of
intense active region blinkers seen by EIS, suggests that the
high-excitation \lam192.9 line is not as sensitive to the events as
the CDS \lam629.7 line.

Based on the EM, it appears that the SMM/UVSP bursts are consistent with the UV bursts.  The most intense of the CDS active region blinkers are  weaker than the IRIS and UVSP events, but likely still qualify as UV bursts.

\newpage

\end{document}